\documentclass[aps,prb,twocolumn,superscriptaddress]{revtex4-2}

\usepackage{tikz}
\usepackage{siunitx}
\usepackage{mathtools}
\DeclarePairedDelimiter\bra{\langle}{\rvert}
\DeclarePairedDelimiter\ket{\lvert}{\rangle}

\begin{document}


\title{Single plasmon transport in one dimensional nanowire}


\author{A. A. D\'iaz-Valles}
\affiliation{Université Bourgogne Europe, CNRS, Laboratoire Interdisciplinaire Carnot de Bourgogne ICB UMR 6303, F-21000 Dijon, France}
\author{B. Rousseaux}
\affiliation{Université Marie et Louis Pasteur, CNRS, Institut FEMTO-ST UMR 6174, F-25000 Besançon, France}
\author{S. Guérin} 
\affiliation{Université Bourgogne Europe, CNRS, Laboratoire Interdisciplinaire Carnot de Bourgogne ICB UMR 6303, F-21000 Dijon, France}
\author{H. Jauslin}
\affiliation{Université Bourgogne Europe, CNRS, Laboratoire Interdisciplinaire Carnot de Bourgogne ICB UMR 6303, F-21000 Dijon, France}
\author{A. Leray} 
\affiliation{Université Bourgogne Europe, CNRS, Laboratoire Interdisciplinaire Carnot de Bourgogne ICB UMR 6303, F-21000 Dijon, France}
\author{G. Colas des Francs}
\email[]{gcolas@ube.fr}
\affiliation{Université Bourgogne Europe, CNRS, Laboratoire Interdisciplinaire Carnot de Bourgogne ICB UMR 6303, F-21000 Dijon, France}



\begin{abstract}
We introduce a unified theoretical framework for single-plasmon transport in one-dimensional nanowires, bridging the quantized electromagnetic Green’s tensor formalism with effective non-Hermitian Hamiltonian models. This approach naturally incorporates propagating surface plasmon polaritons, high-order modes dissipative channels, and intrinsic losses. We investigate both the stationary regime and the spatio-temporal dynamics of a single-plasmon pulse travelling through an atomic chain coupled to a dispersive nanowire. We analyze modal contributions to reflection and transmission spectra for quantum emitter coupled to a silver nanowire, a  configuration proposed as a single-plasmon transistor, and we demonstrate that optimized multi-emitter systems offer significant advantages. In case of one quantum emitter coupled to a silver nanowire at telecom wavelengths, we predict a single-plasmon transmittivity down to 7\% under realistic conditions, and an atomic qubit population of 12\%. Extension to multi-emitter systems using a Löwdin orthogonalization procedure enables a consistent treatment of collective interactions.  We show that optimized positioning with just five emitters enhances plasmon modulation, achieving a transmittivity of 2\% but also reduces coupling losses to one-third compared to the single-emitter case. Our results establish a robust foundation for analyzing and designing plasmonic waveguide quantum electrodynamics systems. 
\end{abstract}


\maketitle

The need for integrated components has stimulated the development of nanophotonic and plasmonic architectures, realizing truly scalable all-optical quantum computing platforms \cite{Lodahl:2015,Maring-Senellart:24,Gonzalez-Tuleda:24}. Optical confinement in nanostructures not only enhances light-matter interaction but also mediates photon-photon interactions—otherwise weak—in direct analogy to cavity quantum electrodynamics (cQED). Waveguide quantum electrodynamics (wQED) exploits evanescent coupling  to create controlled photon-photon quantum correlations for quantum information processing \cite{Turschmann:19,Shemeret:23}. In particular, surface plasmon polaritons (SPPs) guided along plasmonic nanowires present strongly subwavelength confinement, of big interest for integration capabilities and efficient photon-photon mediated interactions \cite{GCF-Barthes-Girard:2016,Marquier-Sauvan-Greffet:2017,SanchezBarquilla-Feist:22}. Quantum emitters coupled to a one dimensional (1D)  plasmon waveguide have been demonstrated to implement single-photon optical transistors \cite{Chang-Sorensen-Demler-Lukin:2007}, two-qubit entanglement \cite{Fleischhauer:2010,GonzalezTuleda-GarciaVidal:2011}, long-range energy transfer \cite{Wenger-GCF:2016}, and nanowire lasers \cite{Bermudez:17}. At the same time, intrinsic material absorption and the presence of multiple radiative and non-radiative channels introduce dissipative effects that significantly influence energy transport \cite{Vest-Greffet:17}. 

Theoretical treatments of wQED generally follow two complementary approaches. Effective Hamiltonian models describe emitter–waveguide interactions using {\it ad hoc} coupling constants and decay rates, allowing compact expressions for reflection and transmission amplitudes \cite{Shen-Fan:05,Chen-Koenderink:11,Liao-Zubairy:15}. In contrast, continuous formulations based on the electromagnetic Green’s tensor provide a description of the electromagnetic environment and its modal structure - although sometimes approximated to a single-mode 1D Green’s function for physical clarity \cite{Fleischhauer:2010,Hummer-Garciavidal:2013,AsenjoGarcia:17}. While both approaches have been extensively employed, their explicit connection and the effect of dissipative high order modes—particularly in lossy plasmonic systems—deserves attention. In this work, we derive effective non-Hermitian Hamiltonians directly from the quantized electromagnetic Green’s tensor of a plasmonic nanowire. This formulation naturally incorporates propagating SPPs, radiative channels, non-radiative decay, and frequency shifts within a unified description. In section \ref{sect:Single}, we first analyze single-plasmon transport for a single emitter coupled to a silver nanowire and quantify the relative contributions of guided and non-guided modes to the reflection and transmission spectra. We then extend the formalism to multi-emitter configurations in section \ref{sect:Multi}. A Löwdin symmetric orthogonalization procedure is employed to define independent polaritonic modes and to obtain a closed set of coupled equations governing collective dynamics. 

\section{Dynamics of the coupled system for a single emitter coupled to a 1D nanowire}
\label{sect:Single}
\vspace*{-4pt} 
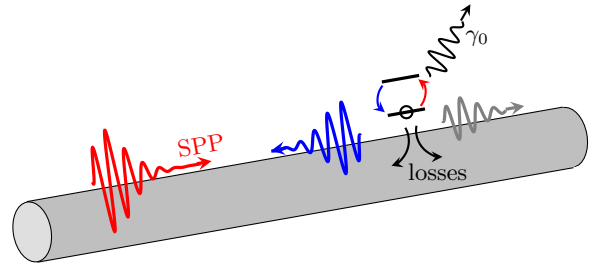
\begin{figure}[h!]
\vspace{-2.75cm}
\centering
 \begin{tikzpicture}[rotate=10,scale=1]
\draw[very thick] (0.3,0.7) --++ (0.5,0) ;
\draw[very thick] (0.3,1.15) --++ (0.5,0) ;
\draw[thick] (0.55,0.7) circle(0.08);
\draw[thick,->, >=stealth,color=red] (0.75,0.75) arc [start angle=-50, end angle=40, radius=0.25];
\draw[thick,->, >=stealth,color=blue] (0.3,1.1) arc [start angle=-50, end angle=40, radius=-0.25];
 \begin{scope}
\clip(2.6,-.4)rectangle(3.2,4);
\filldraw [fill=gray!50,draw=black] (2.6,0) ellipse (0.25cm and 0.4cm);
 \end{scope}
\draw[thick] (-4.6,-0.4) -- (2.6,-0.4);
\draw[thick] (-4.6,0.4) -- (2.6,0.4);
\fill[gray!50](-4.6,-0.4) -- (2.61,-0.4)--(2.61,0.4)--(-4.6,0.4)--cycle;
\filldraw [fill=gray!25,draw=black] (-4.6,0) ellipse (0.25cm and 0.4cm);
\draw[very thick,color=red,xshift=-3cm,yshift=0.5cm,->,>=stealth](0.7,0.01)--++(0.2,0)  node [above,midway,rotate=10]{SPP};
\draw[very thick,color=red,domain=-0.7:0.75,samples=500,xshift=-3.cm,yshift=0.5cm] plot ({\x},{0.05*exp(-5*\x)*(cos(10*\x*3 r)+cos(9*\x*3 r)+cos(11*\x*3 r))});
\draw[very thick,color=blue,domain=-0.5:0.7,samples=500,xshift=-0.8cm,yshift=.5cm] plot ({\x},{-0.035*exp(5*\x)*(cos(10*\x*3 r)+cos(9*\x*3 r)+cos(11*\x*3 r))});
\draw[very thick,color=blue,xshift=-2cm,yshift=0.5cm,<-,>=stealth](0.7,0.01)--++(0.2,0);
\draw[very thick,color=gray,xshift=1.15cm,yshift=0.5cm,->,>=stealth](0.7,0.01)--++(0.25,0);
\draw[very thick,color=gray,domain=-0.7:0.15,samples=500,xshift=1.7cm,yshift=0.5cm] plot ({\x},{0.02*exp(-5*\x)*(cos(10*\x*3 r)+cos(9*\x*3 r)+cos(11*\x*3 r))});
\draw[thick,->, >=stealth,color=black] (0.5,0.5) arc [start angle=20, end angle=-80, radius=0.35]node [below,right,xshift=0.1cm,yshift=-0.08cm]{losses} ;
\draw[thick,->, >=stealth,color=black] (0.65,0.5) arc [start angle=160, end angle=260, radius=0.35];
\begin{scope}[rotate=50]
\draw[thick,color=black,xshift=1.6cm,yshift=0cm,->,>=stealth](0.7,0.01)--++(0.25,0);
\draw[thick,color=black,domain=-0.7:0.15,samples=500,xshift=2.15cm,yshift=0cm] plot ({\x},{0.025*exp(-3.5*\x)*(cos(10*\x*3 r)+cos(9*\x*3 r)+cos(11*\x*3 r))})node [below,xshift=0.2cm]{$\gamma_0$};
\end{scope}
\end{tikzpicture} 
\caption{{Scheme of a two level states emitter coupled to a metallic nanowire and excited by a SPP pulse.}}
\label{fig:scheme}
\end{figure}

\subsection{Lossy plasmon modes quantization and continuous model}

We note $\mathbf{G}({\mathbf r},{\mathbf r}')$ the Green tensor associated to the electric field response at position ${\mathbf r}$ from an excitation localized at ${\mathbf r}'$ in the medium. The electromagnetic field is quantized within the Langevin type model \cite{Knoll-Scheel-Welsch:01} and the electric field operator can be written as
\begin{eqnarray}
\mathbf{\hat{E}}^{+}_\omega(\mathbf{r})=& i\sqrt{\frac{\hbar}{\pi\epsilon_0}}k_0^2
\int d{\mathbf{r}'} \sqrt{\varepsilon''_\omega(\mathbf{r}')}
{\mathbf G}({\mathbf r},{\mathbf r}',\omega)\hat{{\mathbf f}}_\omega({\mathbf r'}) \hspace{0.5cm}
\label{eq:OpE}
\end{eqnarray}
where $k_0=\omega/c$ is the wavenumber, $\varepsilon_\omega=\varepsilon'_\omega+i\varepsilon''_\omega$ refers to the complex dielectric constant and $\hat{{\mathbf f}}_\omega({\mathbf r})$ is the polaritonic vector field operator at position ${\mathbf r}$.  
We follow the procedure described by Castellini {\it et al} \cite{Castellini:18}, but adapted to propagative modes (see appendix \ref{sect:wQEDHamiltonian}) to obtain the
Hamiltonian associated to a single emitter coupled to a plasmonic nanowire, shown schematically on Fig. \ref{fig:scheme}

\begin{eqnarray}
\nonumber 
&&\hat H=\hat H_{\mathrm{QE}}+\hat H_{\mathrm{F}}+\hat H_\mathrm{int} \,,\\
\nonumber 
&&\hat H_{\mathrm{QE}}=\hbar\left(\omega_0-i\frac{\gamma_0}{2}\right)\hat{\sigma}_{+}\hat{\sigma}_{-} \,,\\
\label{hamil}
&&\hat H_{\mathrm{F}}=\sum_{n=0}^\infty \int_0^{+\infty}d\omega \hbar \omega  \left[\hat{\mathbf{a}}_{\omega}^{(n\pm)\dagger}(\mathbf{r}_0)
\hat{\mathbf{a}}_{\omega}^{(n\pm)}(\mathbf{r}_0)\right] \\
\nonumber 
&&\hat H_\mathrm{int}=-i\hbar \hat{\sigma}_{+}\otimes \\
&&\nonumber 
\sum_{n=0}^\infty \int_0^{+\infty}d\omega K_{\omega}^{(n)}({\mathbf r}_0)\left[ \hat{\mathbf{a}}_{\omega}^{(n\pm)}(\mathbf{r}_0) \right]+H.c. \;.
\end{eqnarray}
$\omega_0$ is the transition angular frequency between the ground state $\ket{g}$ and excited state $\ket{e}$ of the emitter and we introduce the coupling operators $\hat{\sigma}_{+}=\ket{e} ~\bra{g}$ and  $\hat{\sigma}_{-}=\ket{g}~\bra{e}$.The first term is the quantum emitter (QE)  energy and we have phenomelogically introduced the decay rate $\gamma_0$ of the excited state\cite{DrezetPRA:17,Dorier:20,Semin2024}. The second term describes  the total energy of the electromagnetic field where $\hat{\mathbf{a}}_{\omega}^{(n)\dagger}(\mathbf{r}_0)$ [$ \hat{\mathbf{a}}_{\omega}^{(n)}(\mathbf{r}_0)$] is the $n^{th}$ polaritonic vector field creation [annihilation] at the position $\mathbf{r}_0$ of the emitter, denoted SPP$_n^\pm$ in the following for simplicity. The $\pm$ signs refer to forward ($+$) and backward ($-$) propagations\cite{notationPM}. The last term describes the emitter-field interaction under the rotating-wave approximation.

We note $\mathbf{d}$ the transition dipole moment of the emitter.  The modal expansion of the Green tensor leads to the following expression of the coupling constant, identical for both forward and backward propagations
\begin{eqnarray}
\vert K_\omega^{(n)}(\mathbf{r}_0)\vert ^2
=\frac{1}{2}\frac{1}{\hbar \pi \epsilon_0}\frac{\omega^2}{c^2} \mathbf{d}\cdot Im \left[{\mathbf G}_{n}({\mathbf r}_0,{\mathbf r}_0,\omega)\right]\cdot  \mathbf{d}^\star 
\label{eq:Kappa2}
 \end{eqnarray} 

The wave function of the hybrid system can be written at time $t$ as
\begin{eqnarray}
\label{wavefun}
\ket{\psi(t)}&=&c_e (t)e^{-i\omega_0 t}\ket{e,\emptyset}
\\
\nonumber
&&
+ \sum_n \int_0^{+\infty}d \omega e^{-i\omega t}\mathbf{c}_\omega^{(n\pm)}(t)\cdot\ket{g,\mathbf{n}_\omega^\pm} 
\end{eqnarray}

$\ket{e,\emptyset}$ refers to the QE in its excited state and no SPP and $\ket{g,\mathbf{n}_\omega^\pm}$ refers to the QE in its ground state and $n^{th}$ forward/backward polariton is excited. 

The dynamics of the wavefunction  is governed by the Hamiltonian (\ref{hamil}), leading to 
\begin{subequations}
\label{Ce_dyn}
\begin{align}
\label{Ce_dyna}
&
\dot{c}_{e}(t)=-\frac{\gamma_0}{2} c_e(t) 
-\int_0^{+\infty}\mathrm{d}\omega \sum_n K_\omega^{(n)} e^{-i(\omega-\omega_0)t} \mathbf{c}_\omega^{(n\pm)}(t)
\\
\label{Ce_dynbc}
&\dot{\mathbf{c}}_\omega^{(n-)}(t)=\dot{\mathbf{c}}_\omega^{(n+)}(t)=K_\omega^{(n)\star} e^{i(\omega-\omega_0)t}c_e(t)
\end{align}
\end{subequations}
so that it presents similar structure as single photon transport discussed in ref. \cite{Greenberg:24} for a single mode waveguide. We extend their procedure to our model for expressing the reflected and the transmitted amplitudes. 
We assume that the system is initially excited by the fundamental guided SPP$_0^+$ and the emitter in its ground state. 
Formal solutions of equations (\ref{Ce_dyn}) can be written as
\begin{subequations}
\label{eq:cn}
\begin{align}
\label{eq:cna}
&\mathbf{c}_\omega^{(0-)}(t)=K_\omega^{(0)\star}\int_0^t  e^{i(\omega-\omega_0)t'}c_e(t')dt' \\
\label{eq:cnb}
&\mathbf{c}_\omega^{(0+)}(t)=\mathbf{c}_\omega^{(0+)}(0)+ K_\omega^{(0)\star} \int_0^te^{i(\omega-\omega_0)t'}c_e(t')dt' \\
\label{eq:cnc}
&\mathbf{c}_\omega^{(n-)}(t)=\mathbf{c}_\omega^{(n+)}(t)=K_\omega^{(n)\star}\int_0^t  e^{i(\omega-\omega_0)t'}c_e(t')dt' 
\end{align}
\end{subequations}
where last line is satifisfied for $n\ge 1$ only. 

\subsection{Stationnary limits}
The stationnary limits of the reflected and transmitted polaritons correspond to the long time limit $t$ \cite{Chen-Koenderink:11,Greenberg:24}
\begin{subequations}
\label{eq:cn_Infty}
\begin{align}
\nonumber
&\mathbf{c}_\omega^{(0-)}(t\rightarrow \infty)=K_\omega^{(0)\star}  \int_0^\infty e^{i(\omega-\omega_0)t'}c_e(t')dt' \\
\label{eq:cn_Inftya}
&\hspace{2cm}=K_\omega^{(0)\star}  C_e(\omega) \\
\label{eq:cn_Inftyb}
&\mathbf{c}_\omega^{(0+)}(t\rightarrow \infty)=\mathbf{c}_\omega^{(0+)}(0)+K_\omega^{(0)\star}  C_e(\omega) \\
\label{eq:cn_Inftyc}
&\mathbf{c}_\omega^{(n-)}(t\rightarrow \infty)=\mathbf{c}_\omega^{(n+)}(t\rightarrow \infty)=K_\omega^{(n)\star} C_e(\omega)
\end{align}
\end{subequations}
{The limit is defined in the sense of distributions, as discussed in appendix \ref{sect:DistribLimit}.} We note that the stationnary limits depends solely on the Laplace transform 
\begin{eqnarray}
C_e(\nu)=\int_0^\infty e^{-i(\omega_0-\nu)t}c_e(t)dt
\end{eqnarray} 
Finally, the spectral response of the coupled system can be solved defining the Laplace transform also for all the other coefficients 
\begin{eqnarray}
C_\omega^{n\pm}(\nu)=\int_0^\infty e^{-i(\omega-\nu)t}c_\omega^{n\pm}(t)dt
\end{eqnarray} 
and from the Laplace transform of the dynamical equations (\ref{Ce_dyn})
\begin{eqnarray}
\label{C_LapA}
&i(\omega_0-\nu)C_e(\nu)=-\frac{\gamma_0}{2} C_e(\nu) 
\\
\nonumber
&\hspace{.5cm} 
-\sum_n \int_0^{+\infty}\mathrm{d}\omega K_\omega^{(n)} \left[C_\omega^{n-}(\nu)
+ C_\omega^{n+}(\nu)\right]
\end{eqnarray}
The n$^{th}$ polariton coefficient can be written as 
\begin{subequations}
\begin{align}
\nonumber
&C_\omega^{n-}(\nu)=C_\omega^{n+}(\nu)=\frac{i}{\nu-\omega_0}K_\omega^{(n)\star} C_e(\nu)\\
&\hspace{1.2cm}=\zeta(\nu-\omega_0) K_\omega^{(n)\star} C_e(\nu)  \;, n\ge 1\\
\nonumber \\
&C_\omega^{0-}(\nu)=\zeta(\nu-\omega) K_\omega^{(0)\star} C_e(\nu) \\
\nonumber \\
&C_\omega^{0+}(\nu)=\zeta(\nu-\omega) \left[c_\omega^{0+}(0)+K_\omega^{(0)\star} C_e(\nu)\right] \\
\nonumber 
&\zeta(\nu-\omega)=\lim_{\epsilon \to 0} \frac{i}{\nu-\omega+i\epsilon}
=\pi \delta(\nu-\omega)+i{\cal P}\left(\frac{1}{\nu-\omega}\right)
\end{align}
\end{subequations}
We introduce the coupling rates and frequency shift due to the coupling to SPP$_n$ (both forward and backward) 
 \begin{eqnarray}
  \nonumber
&& \Gamma_n(\nu)
=4\pi \vert K_\nu^{(n)}\vert^2 \\
 \nonumber
 &&\Omega_n(\nu)= 2{\cal P} \int_0^{+\infty}\mathrm{d}\omega \frac{ \vert K_\omega^{(n)}\vert^2}{\nu-\omega}
 \end{eqnarray}
This leads to the emitter excited state amplitude
\begin{widetext}
\begin{align}
\nonumber
i(\omega_0-\nu)C_e(\nu)&=-\frac{\gamma_0}{2} C_e(\nu) 
-2\sum_{n=0}^\infty \int_0^{+\infty}\mathrm{d}\omega \vert K_\omega^{(n)}\vert^2 C_e(\nu)\zeta(\nu-\omega)
-\int_0^{+\infty}\mathrm{d}\omega K_\omega^{(0)} c_\omega^{(0+)}(0)\zeta(\nu-\omega)\\
C_e(\nu)&=\frac{ \int_0^{+\infty}\mathrm{d}\omega' K_{\omega'}^{(0)}\zeta(\nu-\omega')c_{\omega'}^{(0+)}(0)
}{i(\nu-\omega_0-\sum_n \Omega_n)-\left(\gamma_0/2+\sum_n\Gamma_n/2\right)}\\
\nonumber
 &=\frac{\pi K_{\omega}^{(0)}c_{\omega}^{(0+)}(0)
}{i(\nu-\omega_0-\sum_n \Omega_n)-\left(\gamma_0/2+\sum_n\Gamma_n/2\right)}
+\frac{ i{\displaystyle {\cal P} \int_0^{+\infty}\mathrm{d}\omega' }\dfrac{K_{\omega'}^{(0)}}{\nu-\omega'}c_{\omega'}^{(0+)}(0)
}{i(\nu-\omega_0-\sum_n \Omega_n)-\left(\gamma_0/2+\sum_n\Gamma_n/2\right)}
 \label{eq:Ce_nu}
\end{align}
\end{widetext}

Finally, we express the forward and backward guided mode amplitudes in the long time limit (Eq. \ref{eq:cn_Infty})

\begin{widetext}
\begin{eqnarray}
\nonumber
&&\mathbf{c}_\omega^{(0+)}(\infty)=c_\omega^{0+}(0)\dfrac{\omega-\omega_0-\sum_n \Omega_n+\dfrac{i}{2}\left(\gamma_0+\dfrac{\Gamma_0}{2}+\sum_{n\ge 1}\Gamma_n \right)}{\omega-\omega_0-\sum_n \Omega_n+\dfrac{i}{2}\left(\gamma_0+\sum_n\Gamma_n\right)}
+\frac{K_\omega^{(0)^\star}{\displaystyle {\cal P} \int_0^{+\infty } \mathrm{d}\omega' }\dfrac{K_{\omega'}^{(0)}c_{\omega'}^{(0+)}(0)}{\omega-\omega'}}{\omega-\omega_0-\sum_n \Omega_n+\dfrac{i}{2}\left(\gamma_0+\sum_n\Gamma_n\right)}\\
\nonumber \\ 
\nonumber \\
\label{eq:C0Infty}
&&\mathbf{c}_\omega^{(0-)}(\infty)=\dfrac{-i\dfrac{\Gamma_0}{4}c_\omega^{0+}(0)}{\omega-\omega_0-\sum_n \Omega_n+\dfrac{i}{2}\left(\gamma_0+\sum_n\Gamma_n\right)}
+\frac{K_\omega^{(0)^\star}{\displaystyle {\cal P} \int_0^{+\infty } \mathrm{d}\omega' }\dfrac{K_{\omega'}^{(0)}c_{\omega'}^{(0+)}(0)}{\omega-\omega'}}{\omega-\omega_0-\sum_n \Omega_n+\dfrac{i}{2}\left(\gamma_0+\sum_n\Gamma_n\right)}\\
\nonumber  \\ 
 ~  \nonumber \\
&&\mathbf{c}_\omega^{(n\pm)}(\infty)=\frac{-i \pi K_\omega^{(n)\star} K_\omega^{(0)}c_\omega^{0+}(0)
}{\omega-\omega_0-\sum_n \Omega_n+\dfrac{i}{2}\left(\gamma_0+\sum_n\Gamma_n\right)}
+\frac{ K_\omega^{(n)\star}{\cal P} {\displaystyle \int_0^{+\infty}\mathrm{d}\omega' }\dfrac{K_{\omega'}^{(0)}c_{\omega'}^{(0+)}(0)}{\omega-\omega'}
}{\omega-\omega_0-\sum_n \Omega_n+\dfrac{i}{2}\left(\gamma_0+\sum_n\Gamma_n\right)}
\nonumber
\end{eqnarray}
\end{widetext}

{We can achieve analytical expression considering the following approximation. We assume the coupling strength $K_{\omega'}^{(0)}$ varies slowly over the bandwidth of the initial pulse $c_{\omega'}^{(0+)}$, that is the case for {\it e.g.} a gaussian initial pulse. 
By extending the integration to negative frequencies} 
\begin{eqnarray}
\nonumber
&{\displaystyle {\cal P} \int_0^{+\infty } \mathrm{d}\omega' }\dfrac{K_{\omega'}^{(0)}c_{\omega'}^{(0+)}(0)}{\omega-\omega'}\\
& \rightarrow {\displaystyle {\cal P} \int_{-\infty}^{+\infty } \mathrm{d}\omega' }\dfrac{K_{\omega'}^{(0)}c_{\omega'}^{(0+)}(0)}{\omega-\omega'} =-i\pi K_{\omega}^{(0)}c_{\omega}^{(0+)}
\label{eq:PVapprox}
\end{eqnarray}
where we applied the residue theorem, with a pole at $\omega'=\omega$ .Then, recalling that $\Gamma_0(\omega)=4\pi \vert K_{\omega}^{(0)}\vert^2$, and incorporating implicity the frequency shifts $\Omega_n$ into $\Omega_0 \approx \omega_0+\sum_n \Omega_n(\nu)$, we recover expressions extending reflected and transmitted scattering amplitudes to each modes \cite{Liao-Zubairy:15,Greenberg:24}
 
\begin{eqnarray}
\nonumber
&&\mathbf{c}_\omega^{(0+)}(\infty)\approx c_\omega^{0+}(0)\dfrac{\omega-\Omega_0+\dfrac{i}{2}\left(\gamma_0+\sum_{n\ge 1}\Gamma_n \right)}{\omega-\Omega_0+\dfrac{i}{2}\left(\gamma_0+\sum_n\Gamma_n\right)}
\nonumber \\ 
\nonumber \\
\label{eq:ReflecSingle}
&&\mathbf{c}_\omega^{(0-)}(\infty)\approx \dfrac{-i\dfrac{\Gamma_0}{2}c_\omega^{0+}(0)}{\omega-\Omega_0+\dfrac{i}{2}\left(\gamma_0+\sum_n\Gamma_n\right)}
\\
\nonumber  \\ 
 ~  \nonumber \\
&&\mathbf{c}_\omega^{(n\pm)}(\infty)\approx\frac{-2i \pi K_\omega^{(n)\star} K_\omega^{(0)}c_\omega^{0+}(0)
}{\omega-\Omega_0+\dfrac{i}{2}\left(\gamma_0+\sum_n\Gamma_n\right)}
\nonumber
\end{eqnarray}
  
If we neglect the free space decay rate $\gamma_0$, the reflectivity and transmitivity associated to each mode is given by 
  \begin{eqnarray}
\nonumber
&&T_0(\omega)=\frac{\vert \mathbf{c}_\omega^{(0+)}(\infty)\vert ^2}{\vert c_\omega^{0+}(0)\vert ^2}\approx  \dfrac{(\omega-\Omega_0)^2+\left(\sum_{n\ge 1}\Gamma_n \right)^2/4}{(\omega-\Omega_0)^2+\left(\sum_n\Gamma_n\right)^2/4}
\nonumber \\ 
\nonumber \\
&&R_0=\frac{\vert \mathbf{c}_\omega^{(0-)}(\infty)\vert ^2}{\vert c_\omega^{0+}(0)\vert ^2}\vert ^2\approx \dfrac{\Gamma_0^2/4}{(\omega-\Omega_0)^2+\left(\sum_n\Gamma_n\right)^2/4}
\\
\nonumber  \\ 
 ~  \nonumber \\
&&R_n=\frac{\vert \mathbf{c}_\omega^{(n-)}(\infty)\vert ^2}{\vert c_\omega^{0+}(0)\vert ^2} \approx \frac{\Gamma_0\Gamma_n/4}{(\omega-\Omega_0)^2+\left(\sum_n\Gamma_n\right)^2/4}
\nonumber \\
&&T_n=\frac{\vert \mathbf{c}_\omega^{(n+)}(\infty)\vert ^2}{\vert c_\omega^{0+}(0)\vert ^2} =R_n
\nonumber
\end{eqnarray}
 So,  we observe  that these coefficients satisfy $R_0+T_0+\sum_{n \ge 1}(R_n+T_n)=1$ neglecting free space emission, that is all the emission is coupled to SPP modes.  

\subsection{Numerical simulations}
We consider a silver nanowire of radius 50 nm coupled to a single emitter located at 15 nm from its surface in a configuration  similar to the experimental configuration of ref.\cite{Wenger-GCF:2016}. The dielectric constant of silver is described by Drude model $\epsilon(\omega)=\epsilon_{\infty}-\omega_p^2/(\omega^2+i\gamma_p\omega)$ with $\epsilon_{\infty}=6, ~  \hbar\omega_p=7.9$eV  and $\hbar\gamma_p=51$ meV.

\begin{figure}[h!]
\includegraphics[width=6cm]{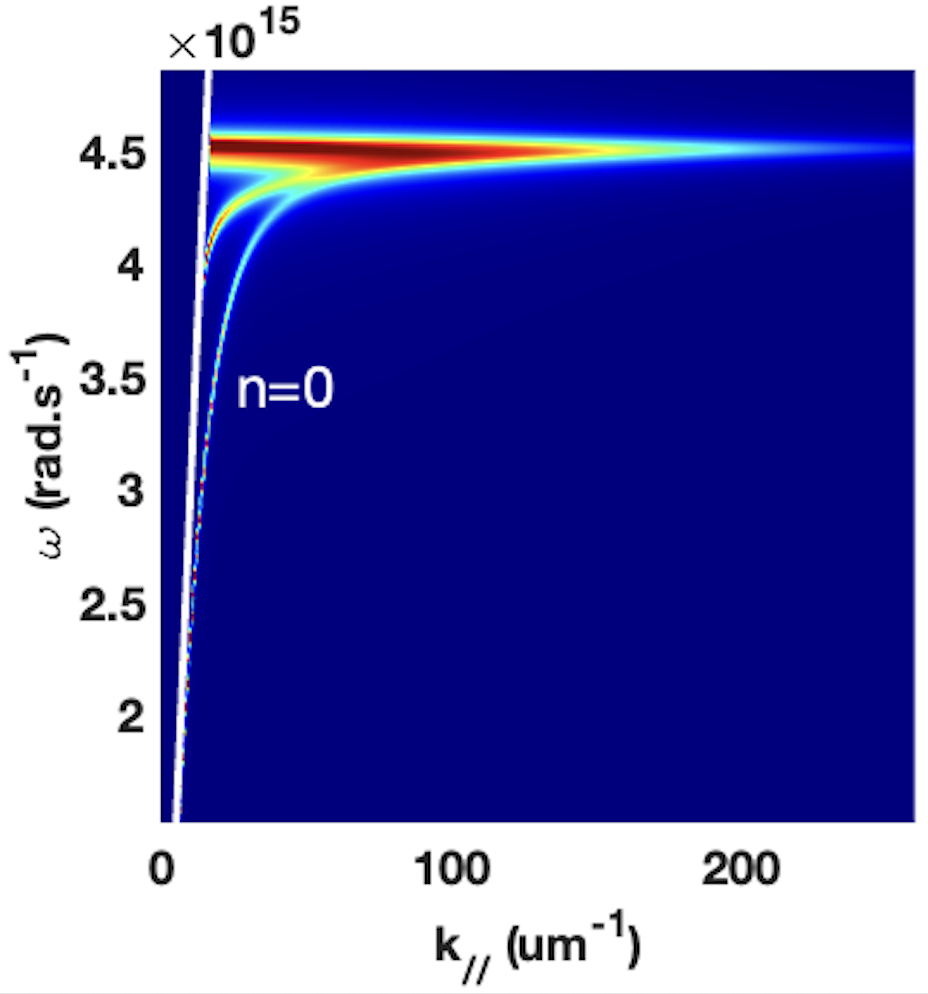}
\caption{Coupling constants (Eq. \ref{eq:Kappa2}) as a function of the parallel wavevector and angular frequency.}
\label{fig:AgNW50nmDisp}
\end{figure}
We represent in Fig. \ref{fig:AgNW50nmDisp} the coupling strength $\vert K _n^\omega\vert^2$. Its maxima follows the dispersion relation of the nanowire SPP modes. We observe no cut-off for the fundamental ($n=0$) mode but a cut-off for other modes (at $\omega_{c1}\approx \SI{4e15}{\radian \cdot s^{-1}}$ for $n=1$). 
In the following, we assume an emission at the telecom wavelength $\lambda=\SI{1.5}{\micro \meter}$ ($\omega_0=\SI{1.3e15}{\radian \cdot s^{-1}}$), for which the nanowire supports a single mode with an effective index $n_{spp}=1.15$ and a propagation length $L_{spp}=\SI{14}{\micro \meter}$. Here, only the $n=0$ mode is a SPP propagating along the nanowire. However, we keep the terminology of  high order modes for $n\ge 1$, although they are not propagative SPPs, but correspond to either radiative or non radiative emission channels (see {\it e.g.} \cite{Barthes-GCF-Bouhelier-Weeber-Dereux:2011}). For larger multimodal nanowires, higher order modes may correspond to propagating guided modes.
  
\begin{figure}[h!]
\centering
\includegraphics[width=6cm]{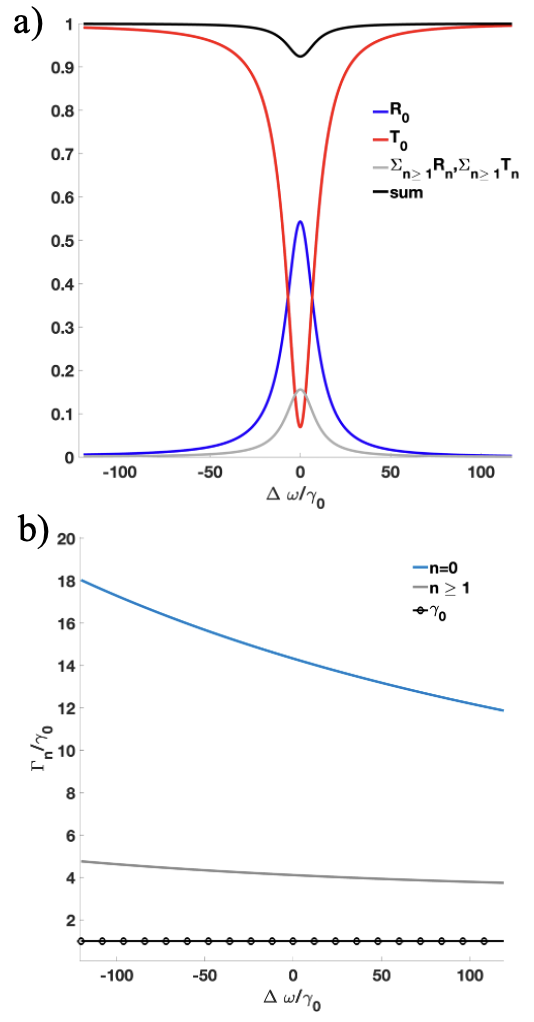}
\caption{a) Modal reflectivity $R_0(\omega)=\vert \mathbf{c}_\omega^{(0-)}(\infty)\vert ^2/\vert c_\omega^{0+}(0)\vert ^2$ and transmittivity $T_0(\omega)=\vert \mathbf{c}_\omega^{(0+)}(\infty)/\vert c_\omega^{0+}(0)\vert ^2$ associated to the fundamental mode and losses $\sum_{n\ge 1} R_n =\sum_{n\ge 1} T_n=\sum_{n\ge 1} \vert \mathbf{c}_\omega^{(n\pm)}(\infty)\vert ^2/\vert c_\omega^{0+}(0)\vert ^2$ associated to high order modes  (Eq. \ref{eq:ReflecSingle}). b) Normalized modal decay rates (Purcell factors).  The emitter free space decay rate is $\gamma_0=1$ meV, $\Omega_0\approx \omega_0$.} 
\label{fig:AgNW50nmSingleReflec}
\end{figure}

The various contributions to the reflectivity and transmitivity are presented in Fig. \ref{fig:AgNW50nmSingleReflec}a). The main contributions correspond to the fundamental mode that was excited but non negligible contributions arise from high orders modes. {The only propagative mode being the fundamental SPP$_0$, solely $R_0$ and $T_0$ measure reflectivity and transmissivity whereas $R_n$ and $T_n$ are associated to high order mode losses. 
These absorption losses in the metal are summarized by the formula $Q_{abs} =\sum_{n\ge 1} \left(R_n +T_n\right)$.

Single plasmon reflectivity and transmitivity achieve $R_0=54\%$ and $T_0=7\%$ at zero detuning $\omega=\omega_0$. These high reflectivity and low transmitivity occurs due to the efficient emission coupling to the fundamental mode.  Such a low single plasmon transmission motivated the concept of single photon optical transistor \cite{Chang-Sorensen-Demler-Lukin:2007}. More quantitatively, we can estimate the coupling efficiency $\beta$-factor from Fig. \ref{fig:AgNW50nmSingleReflec}b): $\beta \approx\Gamma_0/(\Gamma_0+\Gamma_{other})=14/20=70\%$ where $\Gamma_{other}=\gamma_0+\sum_{n\ge 1}\Gamma_n$ refers to emission channels in free space and high orders modes. Non negligible free-space emission at the rate $\gamma_0$ occurs so that the sum $R_0+T_0+Q_{abs}$ in the waveguide (including high order mode losses $Q_{abs}=28\%$) is less than one at zero detuning (93\%). We compute similar, but slightly lower,  modulation of the single plasmon transport from exact Eq. \eqref{eq:C0Infty} instead of the flat coupling approximation of  Eq. \eqref{eq:ReflecSingle}: $R_0=49\%, T_0=9\%$ and $Q_{abs}=32\%$ at zero detuning (not shown).}

\begin{figure}[h!]
\centering
\includegraphics[width=6cm]{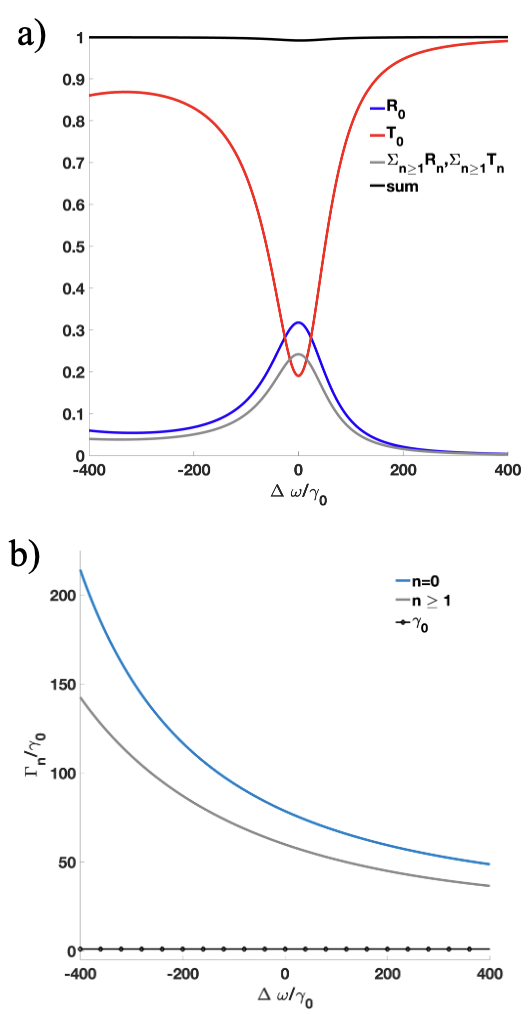}
\caption{Same as Fig. \ref{fig:AgNW50nmSingleReflec} but for a nanowire radius 25 nm and an emitter 5 nm away its surface.}
\label{fig:AgNW25nmSingleReflec}
\end{figure}

We also consider an emitter at 5 nm to a 50 nm diameter silver nanowire on Fig. \ref{fig:AgNW25nmSingleReflec}. 
The modal decay rates significantly vary over the considered spectral range and lead to a Fano asymmetry in the reflectivity and the transmittivity. 

\subsection{{Dynamics of the qubit excitation}}
The dynamics of the qubit amplitude $c_e(t)$ follows 
\begin{eqnarray}
c_e(t)&=&\frac{1}{2\pi}\int_{-\infty}^{\infty}C_e(\nu)e^{i(\omega_0-\nu)t}\mathrm d\nu
\end{eqnarray}
We consider a plasmon pulse propagating from $-x_0$ along the nanowire. We choose $x_0=4.3 v_g/\Delta=\SI{8.6}{\micro \meter}$ before the emitter so that the pulse amplitude is initially negligible at the emitter position ($1\%$ of its maximum). The group velocity of the SPP$_0$ pulse $v_g=0.81c$ is obtained from the dispersion curve Fig. \ref{fig:AgNW50nmDisp}a).  The pulse amplitude at the position of the emitter at time $t=0$ is expressed as   
\begin{eqnarray}
c_{\omega}^{(0+)}(0)=\left(\frac{2}{\pi \Delta^2}\right)^{1/4}e^{-\frac{(\omega-\omega_0)^2}{\Delta^2}}e^{i\tilde k_{spp}(\omega)x_0}
\end{eqnarray}
with the complex propagation contant ${\tilde k_{spp}}(\omega)= n_{spp}(\omega)\omega/c+i/2L_{spp}(\omega)$, that takes into account propagation losses and dispersion. The dispersion remains limited over the considered frequency range: the effective index $n_{spp}$ varies from 1.137 to 1.164 and the propagation length $L_{spp}$ varies from $\SI{15.7}{\micro\meter}$ to $\SI{12.2}{\micro\meter}$.

\begin{figure}[h!]
\includegraphics[width=7cm]{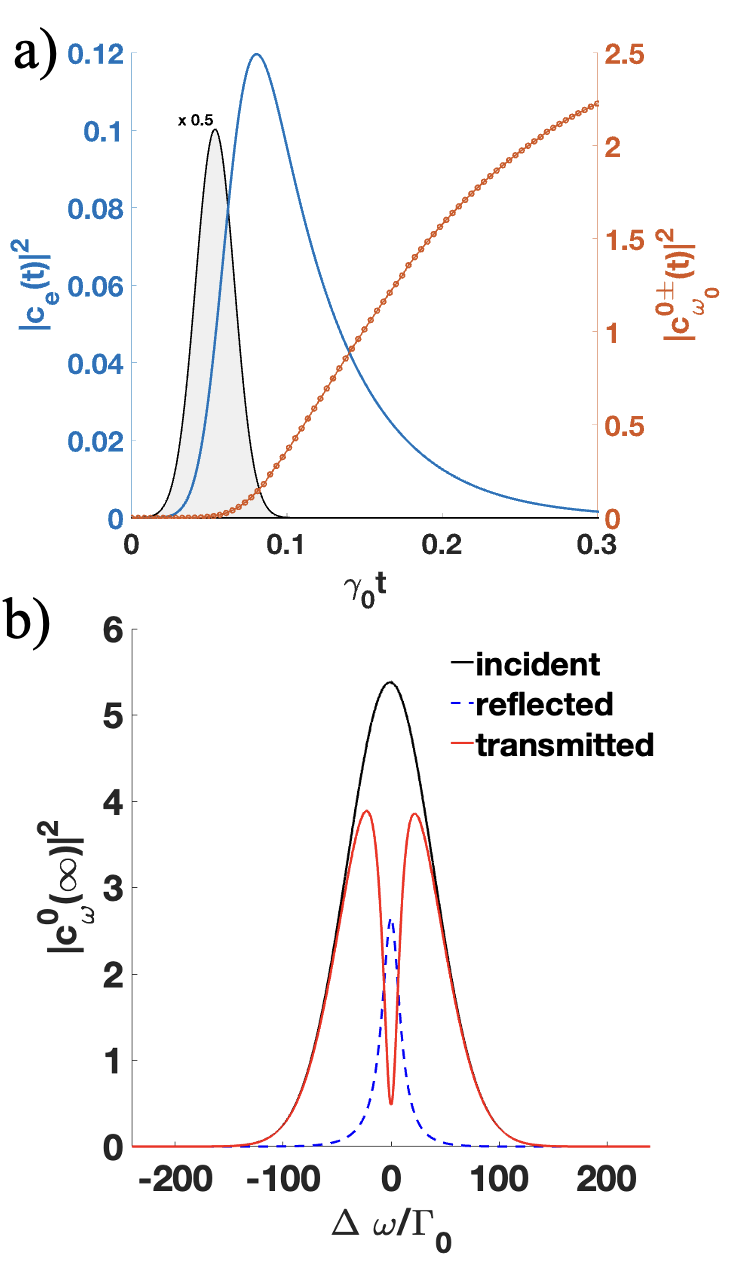}
\caption{a) Time dependance of the atomic $\vert c_e(t)\vert^2$ and SPP$_0$ (at $\omega=\omega_0$) excitations. The gray area refers to the incident pulse enveloppe at the emitter position.b) Spectra of the incident, reflected and transmitted SPP$_0$  at the position of the emitter. The parameters are the same as for Fig. \ref{fig:AgNW50nmSingleReflec}. The incoming SPP$_0$ pulse width is fixed to $\Delta=0.1\omega_0$. $\gamma_0$ refers to the emitter decay rate in free space and $\Gamma_0$ is the coupling strength to the guided mode SPP$_0$.}
\label{fig:PopulationDynSingle}
\end{figure}

The dynamics of excitation of the atomic qubit is plot on Fig.\ref{fig:PopulationDynSingle}a). Propagation losses limits the maximum achievable but a probability around 12\% is obtained for incident SPP$_0$ generated $\SI{8.6}{\micro\meter}$ before the emitter.  Relaxation into the the guided SPP$_0$ is presented on the same figure (only at the emission frequency $\omega=\omega_0$ for clarity). The spectra of the incident, reflected and transmitted guided SPP$_0$ are shown on Fig.\ref{fig:PopulationDynSingle}b) in the stationnary limit, that is the emitter has relaxed into its ground state. We observe a narrow spectrum for the reflected mode, since its width is governed by the qubit decay rate $\gamma_0$ smaller than the incident pulse width $\Delta$. The transmitted SPP$_0$ is modified accordingly compared to the incoming pulse. 

\begin{figure}[h!]
\includegraphics[width=7cm]{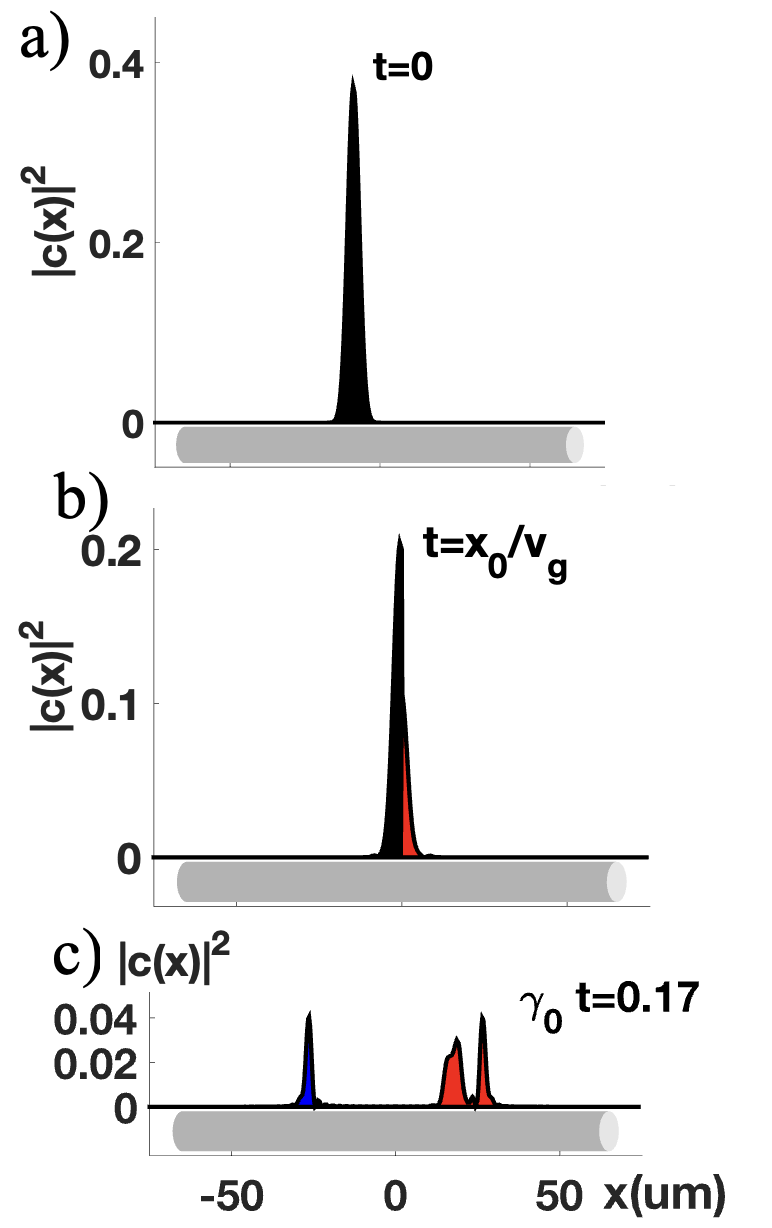}
\caption{Single plasmon pulse propagation at different times. a) Initial position. b) The pulse arrives on the emitter. c) After the interaction. Note that the scales are different since propagation losses occur. Incident pulse in black, transmitted and reflected pulses are in red and blue, respectively.}
\label{fig:SPPpulsePropa1D}
\end{figure}

The reflected and transmitted single plasmon pulse shape obey
\begin{eqnarray}
c_{spp}^\pm(x,t)=\int_{-\infty}^\infty c_\omega^{0\pm}(t-\frac{x}{v_g})e^{\pm i\tilde k_{spp}(\omega)x}e^{-i(\omega-\omega_0)t} \mathrm d\omega \hspace{0.55cm}
\end{eqnarray}
 where $c_\omega^{0\pm}(t)$ are the amplitude of the plasmon excited at the emitter position, given by Eq.(\ref{eq:cna},\ref{eq:cnb}).
The single plasmon pulse shape is represented on Fig. \ref{fig:SPPpulsePropa1D} at the initial time (a), when the incident pulse meets the emitter (b) and at longer time (c). We clearly see the reflected and transmitted pulses, with a characteristic dip in the transmitted pulse, originating from interferences between the incident pulse and surface plasmon coupled emission. This leads to the strong SPP modulation as discussed above. A movie is available in the supplementary materials. It is worth noticing that propagation losses jeopardize long range transport but the strongly subwavelength confinement of the plasmon also ensures important integration capabilities for on chip active control of single plasmon transport. A key strategy is to interface it with a photonic circuitry into a hybrid photonic-plasmonic platform to take benefit of both efficient plasmon-mediated photon-photon interaction and long-range photonic transport \cite{Chang-Sorensen-Demler-Lukin:2007,Lodahl:2015,Weeber-Dubertret:2016}.

\section{Multi-emitter single plasmon transport}
\label{sect:Multi}
For $N_e$ identical emitters, we can generalize the preceding formalism by defining polariton modes excited by each emitter  \cite{Castellini:18,Feist:Nanopht22}. However, care has to be taken in defining a set of orthornormalized polaritonic modes. Applying an orthogonalization procedure, one defines $N_{ind} \le N_e$ independent polaritonic operators $\mathbf{\hat{b}}_{j,\omega}^{(n\pm)}$ associated to forward (+) or backward (-) modes (see appendix \ref{sect:H_Ne}).
\begin{figure}[h!]
\vspace{-3cm}
\centering
 \begin{tikzpicture}[rotate=10,scale=1]
\draw[very thick] (-1.25,0.6) --++ (0.5,0);
\draw[very thick] (-1.25,1.05) --++ (0.5,0) ;
\draw[thick] (-1,0.6) circle(0.08);
\draw[very thick] (0.,0.6) --++ (0.5,0) ;
\draw[very thick] (0.,1.05) --++ (0.5,0) ;
\draw[thick] (0.25,0.6) circle(0.08);
\draw[very thick,dashed] (1.,0.6) --++ (0.5,0) ;
\draw[very thick,dashed] (1.,1.05) --++ (0.5,0) ;
\draw[thick] (1.25,0.6) circle(0.08);
\draw[thick,<-, >=stealth,color=blue,xshift=-1.5cm] (0.75,0.65) arc [start angle=-50, end angle=40, radius=0.25];
\draw[thick,<-, >=stealth,color=red,xshift=-1.5cm] (0.3,1.) arc [start angle=-50, end angle=40, radius=-0.25];
 \begin{scope}
\clip(3.6,-.4)rectangle(4.2,4);
\filldraw [fill=gray!50,draw=black] (3.6,0) ellipse (0.25cm and 0.4cm);
 \end{scope}
\draw[thick] (-4.6,-0.4) -- (3.6,-0.4);
\draw[thick] (-4.6,0.4) -- (3.6,0.4);
\fill[gray!50](-4.6,-0.4) -- (3.61,-0.4)--(3.61,0.4)--(-4.6,0.4)--cycle;
\filldraw [fill=gray!25,draw=black] (-4.6,0) ellipse (0.25cm and 0.4cm);
\draw[very thick,color=red,xshift=-3cm,yshift=0.5cm,->,>=stealth](0.7,0.01)--++(0.2,0)  node [above,midway,rotate=10]{SPP};
\draw[very thick,color=red,domain=-0.7:0.75,samples=500,xshift=-3.cm,yshift=0.5cm] plot ({\x},{0.05*exp(-5*\x)*(cos(10*\x*3 r)+cos(9*\x*3 r)+cos(11*\x*3 r))});
\draw[very thick,color=gray,xshift=1.95cm,yshift=0.5cm,->,>=stealth](0.7,0.01)--++(0.25,0);
\draw[very thick,color=gray,domain=-0.7:0.15,samples=500,xshift=2.5cm,yshift=0.5cm] plot ({\x},{0.02*exp(-5*\x)*(cos(10*\x*3 r)+cos(9*\x*3 r)+cos(11*\x*3 r))});
\end{tikzpicture} 
\caption{Scheme of several two level states emitters coupled to a metallic nanowire and excited by a SPP pulse.}
\label{fig:schemeNe}
\end{figure}
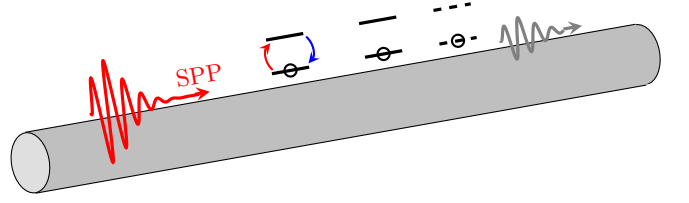

Again, the coupling strengths $K_{n\pm,\omega}^{(i,j)}$ of emitter $i$ to  the polariton $\mathbf{\hat{b}}_{j,\omega}^{(n\pm)}$  are expressed as a function of the Green tensor but also depend on the orthogonalization procedure. We derive analytical expressions of $K_{n\pm,\omega}^{(i,j)}$ as well as initial conditions $\mathbf{c}_{j,\omega}^{(n\pm)}(0)$ for the Löwdin symmetric orthogonalization in the appendix \ref{sect:H_Ne}). 
The wavefunction writes 
\begin{eqnarray}
\label{wavefunNe}
\ket{\psi(t)}&=&\sum_{i=1}^{N_e}c_e^{(j)} (t)e^{-i\omega_0 t}\ket{e^i,\emptyset}
\\
\nonumber
&&
+ \sum_n \sum_{j=1}^{N_{ind}}\int_0^{+\infty}d \omega e^{-i\omega t}\mathbf{c}_{j,\omega}^{(n\pm)}(t)\cdot\ket{g,\mathbf{n}_{j,\omega}^{\pm}} 
\end{eqnarray}

At this point, the procedure is very similar to the single emitter configuration and is developed in the appendix. For a system initially excited by the fundamental guided SPP$_0^+$ and all emitters in their ground state, the emitter's coefficients are solutions of the system of equations 
\begin{widetext}
\begin{eqnarray}
\underbrace{
\begin{pmatrix}
W_{11} & W_{12}& \ldots & W_{1N_e} \\
\vdots & \vdots  & \ddots & \vdots \\
W_{N_e 1} &W_{N_e2} &\ldots &  W_{N_e N_e}
\end{pmatrix}
}_{(N_e \times N_e)}
\begin{pmatrix}
C_e^{(1)}(\omega) \\
 \vdots   \\
C_e^{(N_e)}(\omega) \\
\end{pmatrix}
=2\pi 
\underbrace{
\begin{pmatrix}
K_{0+,\omega}^{(1,1)}
& \ldots & K_{0+,\omega}^{(1N_{ind})} \\
\vdots  & \ddots & \vdots \\
K_{0+,\omega}^{(N_{e},1)} &\ldots &  K_{0+,\omega}^{(N_e,N_{ind})}
\end{pmatrix}
}_{(N_e\times N_{ind})}
\begin{pmatrix}
  c_{1,\omega}^{0+}(0) \\
 \vdots   \\
  c_{N_{ind},\omega}^{0+}(0) \\
\end{pmatrix}
\label{eq:NemMatrix}
\end{eqnarray}
\end{widetext}
with the matrix characterizing the coupled system 
\begin{eqnarray}
\nonumber
&&W_{ii}=i\left(\omega-\omega_0-\sum_n  \sum_{j=1}^{N_{ind}} \Omega_n^{(i,j)} \right)-\frac{\gamma_0}{2} -\sum_n  \sum_{j=1}^{N_{ind}} \frac{\Gamma_n^{(i,j)}}{2}
\\
&&W_{ik}=-\pi \sum_n   \sum_{j=1}^{N_{ind}}  \left[K_{n-,\nu}^{(i,j)}  K_{n-,\nu}^{(k,j)\star}+K_{n+,\nu}^{(i,j)}  K_{n+,\nu}^{(k,j)\star}\right]  \\
\nonumber
&&-i \sum_n   \sum_{j=1}^{N_{ind}}\int_0^{+\infty}\mathrm{d \omega'}\frac{\left[K_{n-,\omega'}^{(i,j)}  K_{n-,\omega'}^{(k,j)\star}+K_{n+,\omega'}^{(i,j)}  K_{n+,\omega'}^{(k,j)\star}\right] }{\omega'-\omega} 
\\
\nonumber
&&\Omega_n^{(i,j)} = \int_0^{+\infty}\mathrm{d \omega}\frac{\vert K_{n-,\omega}^{(i,j)}\vert^2}{\omega-\nu} + \int_0^{+\infty}\mathrm{d \omega}\frac{\vert K_{n+,\omega}^{(i,j)}\vert^2}{\omega-\nu} \\
&&\Gamma_n^{(i,j)} = 2\pi \vert K_{n-,\omega}^{(i,j)}\vert^2+2\pi \vert K_{n+,\omega}^{(i,j)}\vert^2
\nonumber
\end{eqnarray}

Finally, the reflectivity and the transmittivity are given by 
  \begin{eqnarray}
\nonumber
&&T_0(\omega)=\frac{\sum_{i=1}^{N_e}\vert \mathbf{c}_{i,\omega}^{(0+)}(\infty)\vert ^2}{\sum_{i=1}^{N_e}\vert  \mathbf{c}_{i,\omega}^{0+}(0)\vert ^2}\\
\nonumber \\ 
\nonumber \\
&&R_0(\omega)=\frac{\sum_{i=1}^{N_e}\vert \mathbf{c}_{i,\omega}^{(0-)}(\infty)\vert ^2}{\sum_{i=1}^{N_e}\vert  \mathbf{c}_{i,\omega}^{0+}(0)\vert ^2}
\end{eqnarray}
and for higher order modes, corresponding to coupling losses here,   
 \begin{eqnarray}
&&R_n(\omega)=\frac{\sum_{i=1}^{N_e}\vert \mathbf{c}_{i,\omega}^{(n-)}(\infty)\vert ^2}{\sum_{i=1}^{N_e}\vert  \mathbf{c}_{i,\omega}^{0+}(0)\vert ^2}
\nonumber \\
&&T_n(\omega)=\frac{\sum_{i=1}^{N_e}\vert \mathbf{c}_{i,\omega}^{(n+)}(\infty)\vert ^2}{\sum_{i=1}^{N_e}\vert  \mathbf{c}_{i,\omega}^{0+}(0)\vert ^2}
\nonumber
\end{eqnarray}

\begin{figure}[h!]
\includegraphics[width=8cm]{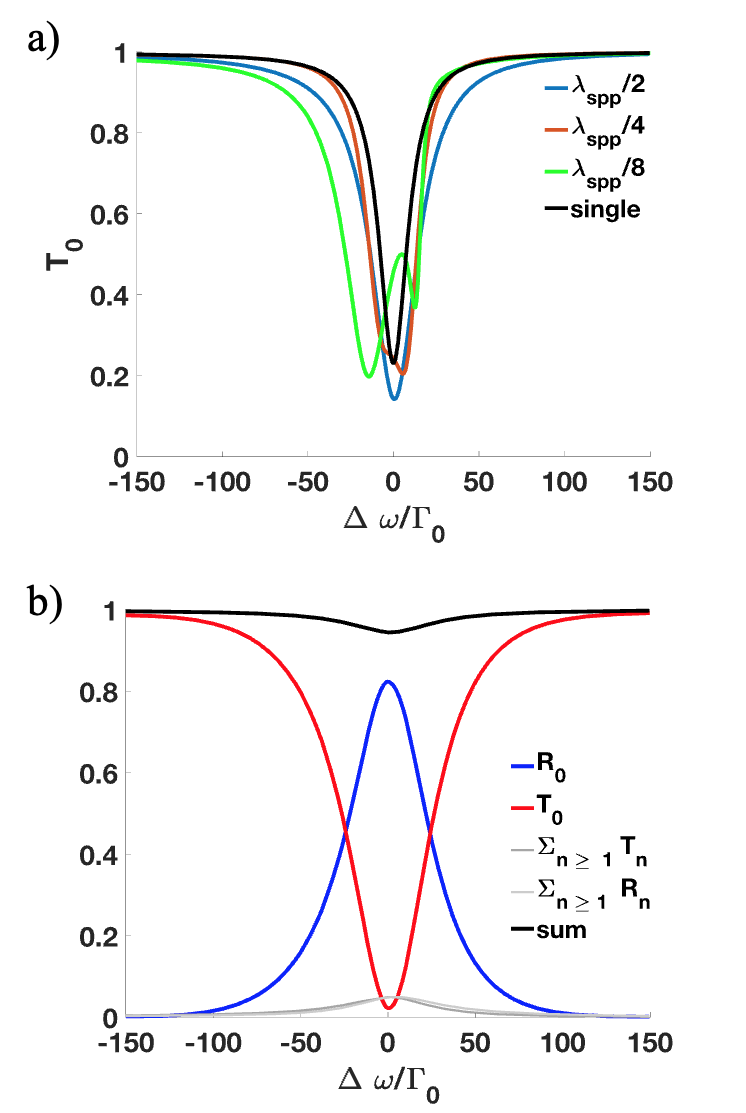}
\caption{a) Transmittivity spectra for two emitters coupled to the nanowire. The distance between the two emitters varies from $\lambda_{spp}/8$ to $\lambda_{spp}/2$. The single emitter case is reproduced for comparison purposes. b) Modal reflectivities and transmittivities for 5 emitters close to the nanowire and separated from each other by $\lambda_{spp}/2$.} 
\label{fig:SpecMultiEmitter.eps}
\end{figure}
We report on Fig. \ref{fig:SpecMultiEmitter.eps}a) the total transmittivity for two emitters considering various separation distances. For small separation, we observe a strong deformation of the transmission compared to the single emitter case. 
However,  for $k_{spp}\Delta z=\pi$ ($\Delta z= \lambda_{spp}/2$), the second emitter is at the antinode of the mode excited by the first one so that the  effects cumulate, and we observe a smaller transmittivity for this optimized configuration. Additional emitters are considered in Fig. \ref{fig:SpecMultiEmitter.eps}b) that accentuates again the transmittivity dip and width ($R_0=66\%$ and $T_0=4\%$ for 2 emitters, $R_0=86\%$ and $T_0=2\%$ for 5  emitters). High order modes losses reduces to $Q_{abs}=10\%$ for 5 emitters, that is one third the losses observed for a single emitter. Thus, the collective interaction of multiple emitters with a single plasmon enhances control over plasmon transport by both increasing signal modulation and reducing coupling losses. We have considered an array of quantum emitters aligned along the nanowire but more complex configurations, {\it e.g} ring of emitters could be of interest to further decreases coupling and free-space losses \cite{Brion:25}.

\begin{figure}[h!]
\includegraphics[width=8cm]{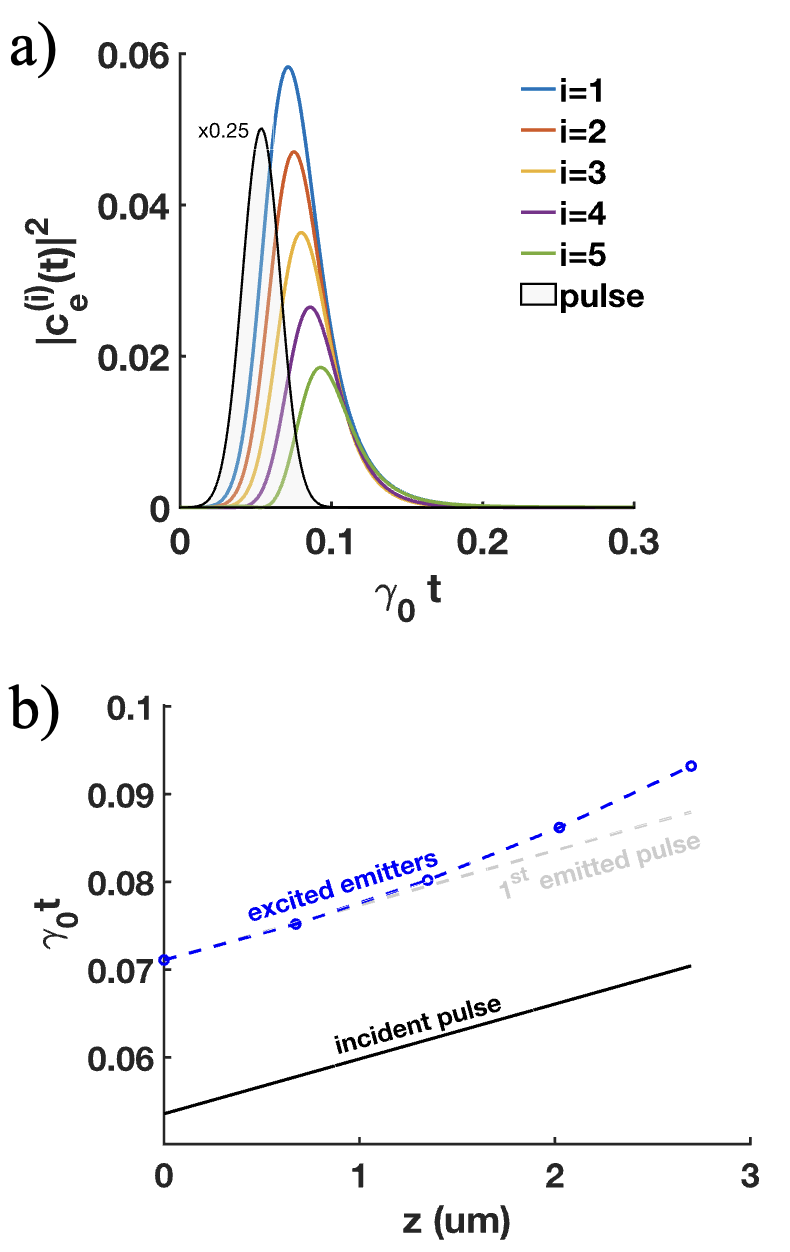}
\caption{a) Dynamics of excited state populations for the five emitters. b) Characteristics times of the single plasmon transport. Incident pulse refers to time at which incident pulse arrives at position z in absence of emitters. 1$^{st}$ emitted pulse corresponds to the propagation of the pulse emitted by the first emitter in absence of the others emitters. Excited emitters indicate the time at which the five emitters are successively excited.   All the parameters are identical to Fig. \ref{fig:SpecMultiEmitter.eps}.}
\label{fig:PopulationDynNe5}
\end{figure}

We also plot on Fig. \ref{fig:PopulationDynNe5}a)  the population dynamics of the five emitters. We observe successive excitation of the five emitters. Figure \ref{fig:PopulationDynNe5}b) represents some characterics times of the single transport dynamics. We compare the times at which the emitters are excited to the temporal propagation of pulses guided along a bare nanowire. We observe a delay between the arrival of the pulse and the excitation of each emitter. This delay accumulates along the chain of emitters, revealing the complex dynamics of single plasmon transport. 

\section{Conclusion}
We have developed a Green-function-based effective Hamiltonian for single-plasmon transport in nanowires unifying continuous electromagnetic field quantization with discrete non-Hermitian models of wQED. By deriving coupling strengths, decay rates, and frequency shifts directly from the modal Green tensor, this formalism self-consistently incorporates propagating surface plasmon polaritons, higher-order emission channels, and intrinsic material losses. Applied to a silver nanowire coupled to a quantum emitter, our approach demonstrates a control of single-plasmon transmittivity down to 7\% at telecom wavelength, despite dissipative effects. The modal analysis clarifies the distinct roles of guided and non-guided contributions in shaping transport properties and coupling losses.
The extension to multiple emitters 
enables a unified description of collective interactions.  Chain of atoms enables further controlled modulation of energy transport, achieving for instance 82\% reflectivity and 2\% transmittivity for five emitters, while simultaneously reducing coupling losses. Additionnally, we characterize the spatio-temporal propagation of along the atomic chain, revealing the complex dynamics of the emitters-single plasmon interaction and pulse deformation along a dispersive nanowire. This work establishes a robust foundation for analyzing plasmonic wQED systems and may facilitate the quantitative design of integrated quantum nanophotonic devices. A promising future direction is its integration with photonic circuitry, creating hybrid photonic-plasmonic  platforms that leverage both efficient plasmon-mediated photon-photon interactions and long-range photonic transport.

\appendix
\section{wQED Hamiltonian}
\label{sect:wQEDHamiltonian}


The Hamiltonian of a single emitter coupled to the plasmonic nanowire is expressed as
\begin{eqnarray}
\label{hamil2}
&&\hat H=\hat H_{\mathrm{at}}+\hat H_{\mathrm{F}}+\hat H_\mathrm{int} \,,\\
\nonumber 
&&\hat H_{\mathrm{at}}=\hbar\left(\omega_0-i\frac{\gamma_0}{2}\right)\hat{\sigma}_{+}\hat{\sigma}_{-} \,,\\
\nonumber 
&&\hat H_{\mathrm{F}}=\int d\mathbf{r} \int_0^{+\infty}d\omega\ \hbar\omega \hat{\mathbf{f}}_\omega^{\dagger}(\mathbf{r})\cdot\hat{\mathbf{f}}_\omega(\mathbf{r}) \,,\\
\nonumber 
&&\hat H_\mathrm{int}=-\left[\hat{\sigma}_{+}\otimes \int_0^{+\infty}d\omega \mathbf{d}\cdot\hat{\mathbf{E}}^{+}_\omega(\mathbf{r}_0)+H.c.\right] \;.
\end{eqnarray}
$\omega_0$ is the transition angular frequency between the ground state $\ket{g}$ and excited state $\ket{e}$ of the emitter and we introduce the coupling operator $\hat{\sigma}_{+}=\ket{e} ~\bra{g}$ and  $\hat{\sigma}_{-}=\ket{g}~\bra{e}$. In equation (\ref{hamil2}), the first term is the QE energy and we have phenomelogically introduced the decay rate $\gamma_0$ of the excited state. The second term describes  the total energy of the electromagnetic field where $\hat{\mathbf{f}}^{\dagger}({\mathbf r})$ ($\hat{\mathbf{f}}({\mathbf r})$) is the SPP polaritonic vector field operator at the position ${\mathbf r}$ associated to the creation (annihilation) of a quantum in the presence of the nanowire. The last term describes the emitters-field interaction under the rotating-wave approximation.

$\hat{\mathbf{E}}^{+}_\omega$ is the electric field operator associated to the field in presence of the nanowire. We note $\mathbf{G}({\mathbf r},{\mathbf r}')$ the Green tensor associated to the electric field response at position ${\mathbf r}$ from an excitation localized at ${\mathbf r}'$ in the medium. The electromagnetic field is quantized within the Langevin type model \cite{Knoll-Scheel-Welsch:01} and the electric field operator can be written as
\begin{eqnarray}
\mathbf{\hat{E}}^{+}_\omega(\mathbf{r})=& i\sqrt{\frac{\hbar}{\pi\epsilon_0}}k_0^2
\int d{\mathbf{r}'} \sqrt{\varepsilon''_\omega(\mathbf{r}')}
{\mathbf G}({\mathbf r},{\mathbf r}',\omega)\hat{{\mathbf f}}_\omega({\mathbf r'}) \hspace{0.5cm}
\label{eq:OpE}
\end{eqnarray}
where $k_0=\omega/c$ is the wavenumber.  

\subsection{Mode selective quantization}
The Green's tensor associated to the nanowire is written as a sum over the SPP modes
\begin{eqnarray}
{\mathbf G}({\mathbf r},{\mathbf r}',\omega)=\sum_{n=0}^\infty {\mathbf G}_n({\mathbf r},{\mathbf r}',\omega)
\end{eqnarray}
where the mode expansion is written as a function of cylindrical harmonics \cite{tai93}.
 
The interaction Hamiltonian is \cite{Castellini:18}
 \begin{eqnarray}
\nonumber
&\hat H_\mathrm{int}=-i\hbar \left[\hat{\sigma}_{+}\otimes \int_0^{+\infty}d\omega  \int d\mathbf{r} g_{\omega,n}({\mathbf r}_0,{\mathbf r})\cdot\hat{f}_\omega(\mathbf{r}_0)\right] +H.c.\;,\\
 \label{eq:Hint}
&g_{\omega}^{(n)}({\mathbf r}_0,{\mathbf r})=\sqrt{\frac{1}{\hbar \pi \epsilon_0}}\frac{\omega^2}{c^2}  \sqrt{\varepsilon''_\omega(\mathbf{r})}
\mathbf{d}\cdot\ {\mathbf G}_{n}({\mathbf r}_0,{\mathbf r},\omega)
\end{eqnarray}
The cylindrical harmonics satisfy the orthogonality relation such that \cite{tai93}
\begin{eqnarray}
 \int d\mathbf{r} g^\star _{\omega,n}({\mathbf r}_0,{\mathbf r}) \cdot g_{\omega',n'}({\mathbf r}_0,{\mathbf r}) =0 \;, \; \text{for} n \ne n' 
 \end{eqnarray}

and we can introduce the $n^{th}$ bosonic operators 
\begin{subequations}
\begin{align}
&&\hat a_{\omega}^{(n)}({\mathbf r}_0)= \frac{1}{\kappa_\omega^{(n)}(\mathbf{r}_0)} \int d\mathbf{r} g_{\omega}^{(n)}({\mathbf r}_0,{\mathbf r}) \cdot \hat{{\mathbf f}}_\omega({\mathbf r}) \\
&& \vert \kappa_\omega^{(n)}(\mathbf{r}_0)\vert ^2= \int d\mathbf{r} g _{\omega,n}({\mathbf r}_0,{\mathbf r}) \cdot g^\star_{\omega,n}({\mathbf r}_0,{\mathbf r})\\
&=&\frac{1}{\hbar \pi \epsilon_0}\frac{\omega^2}{c^2} \mathbf{d}\cdot Im \left[{\mathbf G}_{n}({\mathbf r}_0,{\mathbf r}_0,\omega)\right]\cdot  \mathbf{d}^\star 
 \end{align} 
\end{subequations}

These operators satisfy the commutation relations by construction \cite{Castellini:18} 
\[\left[\hat a_{\omega}^{(n)},\hat a_{\omega'}^{(n') \dagger} \right]=\delta_{nn'}\delta(\omega-\omega')\]
The interaction Hamiltonian reads
\begin{eqnarray}
\hat H_{int}=-i\hbar \hat{\sigma}_{+}\otimes \int_0^{+\infty}d\omega \sum_n\kappa_{\omega}^{(n)}({\mathbf r}_0)\hat a_{\omega}^{(n)}(\mathbf{r}_0)+H.c.\hspace{1cm}
\end{eqnarray}
Finally, omitting dark modes not populated by the interaction with the emitter, the Hamiltonian simplifies to \cite{Castellini:18}
\begin{eqnarray}
 \nonumber
&&\hat H=\hbar\left(\omega_0-i\frac{\gamma_0}{2}\right)\hat{\sigma}_{+}\hat{\sigma}_{-} 
+\int_0^{+\infty}d\omega \hbar \omega \sum_n \hat a_{\omega}^{(n)\dagger}(\mathbf{r}_0)
 \hat a_{\omega}^{(n)}(\mathbf{r}_0)\\
&&-i\hbar \hat{\sigma}_{+}\otimes \int_0^{+\infty}d\omega \sum_n\kappa_{\omega}^{(n)}({\mathbf r}_0)\hat a_{\omega}^{(n)}(\mathbf{r}_0)+H.c.
 \label{eq:Hbright}
\end{eqnarray}
Since we are interested to forward and backward propagations, we introduce the associated operators  through the 2D-Fourier transform of the Green's tensor. 
\begin{eqnarray}
\nonumber
{\mathbf G}_{n}({\mathbf r}_1,{\mathbf r}_2,\omega)&=&\frac{1}{2\pi}\int_{-\infty}^\infty {\mathbf G}_{n}^{2D}({\mathbf r}_{1\parallel},{\mathbf r}_{2\parallel},k_z)e^{ik_z(z_2-z_1)}dk_z\\
\nonumber
&=&\frac{1}{2\pi}\int_{-\infty}^0 {\mathbf G}_{n}^{2D}({\mathbf r}_{1\parallel},{\mathbf r}_{2\parallel},k_z)e^{ik_z(z_2-z_1)}dk_z\\
\nonumber
 && +\frac{1}{2\pi}\int_0^{\infty} {\mathbf G}_{n}^{2D}({\mathbf r}_{1\parallel},{\mathbf r}_{2\parallel},k_z)e^{ik_z(z_2-z_1)}dk_z \\
\nonumber
&=&\underbrace{\frac{1}{2\pi}\int_0^{+\infty}{\mathbf G}_{n}^{2D}({\mathbf r}_{1\parallel},{\mathbf r}_{2\parallel},k_z)e^{-ik_z(z_2-z_1)}dk_z}_{{\mathbf G}_{n}^-({\mathbf r}_1,{\mathbf r}_2,\omega) \;; \text{backward propagation}}\\
\nonumber
 &&  +\underbrace{\frac{1}{2\pi}\int_0^{\infty} {\mathbf G}_{n}^{2D}({\mathbf r}_{1\parallel},{\mathbf r}_{2\parallel},k_z)e^{ik_z(z_2-z_1)}dk_z}_{{\mathbf G}_{n}^+({\mathbf r}_1,{\mathbf r}_2,\omega) \;; \text{forward propagation}} \\
   \nonumber \\
   g_{\omega}^{(n+)}({\mathbf r}_0,{\mathbf r})&=&\sqrt{\frac{1}{\hbar \pi \epsilon_0}}\frac{\omega^2}{c^2}  \sqrt{\varepsilon''_\omega(\mathbf{r})}
\mathbf{d}\cdot\ {\mathbf G}_{n+}({\mathbf r}_0,{\mathbf r},\omega)\\
 g_{\omega}^{(n-)}({\mathbf r}_0,{\mathbf r})&=&\sqrt{\frac{1}{\hbar \pi \epsilon_0}}\frac{\omega^2}{c^2}  \sqrt{\varepsilon''_\omega(\mathbf{r})}
\mathbf{d}\cdot\ {\mathbf G}_{n-}({\mathbf r}_0,{\mathbf r},\omega)
\end{eqnarray}
Forward/backward propagating SPP operators read
\begin{eqnarray}
&&\hat a_{\omega}^{(n+)}({\mathbf r}_0)= \frac{1}{K_\omega^{(n)}(\mathbf{r}_0)} \int d\mathbf{r} g_{\omega}^{(n+)}({\mathbf r}_0,{\mathbf r}) \cdot \hat{{\mathbf f}}_\omega({\mathbf r}) \\
&&\hat a_{\omega}^{(n-)}({\mathbf r}_0)= \frac{1}{K_\omega^{(n)}(\mathbf{r}_0)} \int d\mathbf{r} g_{\omega}^{(n-)}({\mathbf r}_0,{\mathbf r}) \cdot \hat{{\mathbf f}}_\omega({\mathbf r}) \\
&& \vert K_\omega^{(n)}(\mathbf{r}_0)\vert ^2= \frac{1}{2}\vert \kappa_\omega^{(n)}(\mathbf{r}_0)\vert ^2\\
&&=\frac{1}{2\hbar \pi \epsilon_0}\frac{\omega^2}{c^2} \mathbf{d}\cdot Im \left[{\mathbf G}_{n}({\mathbf r}_0,{\mathbf r}_0,\omega)\right]\cdot  \mathbf{d}^\star 
 \end{eqnarray}
 using the reciprocity property  $${\mathbf G}_{n}({\mathbf r}_1,{\mathbf r}_2,\omega)=\left[{\mathbf G}_{n}({\mathbf r}_2,{\mathbf r}_1,\omega)\right]^T$$ so that \\
 $Im \left[{\mathbf G}_{n+}({\mathbf r}_0,{\mathbf r}_0,\omega)\right]=Im \left[{\mathbf G}_{n-}({\mathbf r}_0,{\mathbf r}_0,\omega)\right]=Im \left[{\mathbf G}_{n}({\mathbf r}_0,{\mathbf r}_0,\omega)\right]/2$.

 Then the Hamiltonian  Eq. (\ref{eq:Hbright})  becomes the hamiltonian Eq.  \eqref{hamil} used in the main text.

\section{Many emitters}
\label{sect:H_Ne}
For $N_e$ identical emitters, the Hamiltonian \eqref{hamil} generalizes to 
 \begin{widetext}
 \begin{align}
  \nonumber
\hat H=&\sum_{i=1}^{N_e}\left\{ \hbar\left(\omega_0-i\frac{\gamma_0}{2}\right)\hat{\sigma}_{+}^{(i)}\hat{\sigma}_{-}^{(i)} 
+\int_0^{+\infty}d\omega \hbar \omega \sum_n \hat a_{\omega}^{(n-)\dagger}(\mathbf{r}_i)
 \hat a_{\omega}^{(n-)}(\mathbf{r}_i)+\int_0^{+\infty}d\omega \hbar \omega \sum_n \hat a_{\omega}^{(n+)\dagger}(\mathbf{r}_i)
 \hat a_{\omega}^{(n+)}(\mathbf{r}_i)  \right. \\
 \nonumber
 & \left. -i\hbar \hat{\sigma}_{+}^{(i)}\otimes \int_0^{+\infty}d\omega \sum_nK_{\omega}^{(n)}({\mathbf r}_i)\hat a_{\omega}^{(n-)}(\mathbf{r}_i)
-i\hbar \hat{\sigma}_{+}^{(i)} \otimes \int_0^{+\infty}d\omega \sum_nK_{\omega}^{(n)}({\mathbf r}_0)\hat a_{\omega}^{(n+)}(\mathbf{r}_i)
+H.c. \right\}
\end{align}
\end{widetext}
but the emitter centered operators are not orthogonal {\it a priori} \cite{Castellini:18,Feist:Nanopht22}. Their commutation relations are
\begin{eqnarray}
\left[ \hat a_{\omega}^{(n+)}(\mathbf{r}_i) , \hat a_{\omega}^{(n+)\dagger}(\mathbf{r}_j) \right]=\delta_{nn'}\delta(\omega-\omega')\mu_{\omega,n}^{ij}\\
\nonumber
\mu_{ij}^\pm =\frac{1}{\hbar \pi \epsilon_0}\frac{\omega^2}{c^2}\frac{Im[{\mathbf d^{(i)}}\cdot {\mathbf G}_{n\pm}({\mathbf r}_i,{\mathbf r}_j,\omega) \cdot {\mathbf d^{(j)}}]}{K_{\omega}^{(n)}({\mathbf r}_i)K_{\omega}^{(n)}({\mathbf r}_j)}
\end{eqnarray}
A set of $N_{ind}\le N_e $ individual  orthogonal operators  $\mathbf{\hat b}_{\omega,j}^{(n\pm)}$ are defined following the L{\"owdin} orthonomalization procedure \cite{Castellini:18}. We first express the $(N_e \times N_e)$ overlap matrix 
\begin{eqnarray}
M_{\omega,n\pm}=
\begin{pmatrix}
1 & \mu_{\omega,n\pm}^{2,1}& \ldots & \mu_{\omega,n\pm}^{N_e,1} \\
\mu_{\omega,n\pm}^{1,2} &1& \ldots & \mu_{\omega,n\pm}^{1,N_e} \\
\vdots & \vdots  & \ddots & \vdots \\
\mu_{\omega,n\pm}^{N_e,1} &\mu_{\omega,n\pm}^{N_e,2} &\ldots &  \mu_{\omega,n\pm}^{N_e,N_e}
\end{pmatrix}
\end{eqnarray}
After diagonalization, $ T_{\omega,n\pm}^\dagger M_{\omega,n\pm}T_{\omega,n\pm}=diag(\lambda_{\omega,n\pm}^1, \ldots \lambda_{\omega,n\pm}^{N_{ind}}, 0\ldots 0)$, one can define  L{\"owdin}  operators 
\begin{eqnarray}
\mathbf{\hat b}_{\omega,j}^{(n\pm)} =\frac{1}{\sqrt{\lambda_{\omega,n\pm}^j}}\sum_{i=1}^{N_e} T_{\omega,n\pm}^{ij} \mathbf{\hat a}_{\omega}^{(n\pm)}(\mathbf{r}_i) 
\label{eq:cj0}
\end{eqnarray}
and coupling strength $K_{n\pm,\omega}^{(i,j)} =K_{\omega}^{(n)}({\mathbf r}_i)\sqrt{\lambda_{\omega,n\pm}^j}T_{\omega,n\pm}^{ij\star}$ so that interaction Hamiltonian and the dynamics of the wavefunction follows 
 \begin{widetext}
 \begin{eqnarray} 
&&\hat H_{int}=-i\hbar  \sum_{i=1}^{N_e}\hat{\sigma}_{+}^{(i)}\otimes \int_0^{+\infty}d\omega \sum_n \sum_{j=1}^{N_{ind}}\left[K_{n-,\omega}^{(i,j)}  \mathbf{ \hat b}_{\omega,j}^{n-} +K_{n+,\omega}^{(i,j)} \mathbf{ \hat b}_{\omega,j}^{n+}\right] 
+H.c.
\\
\label{eq:Ce_dynNe}
&&\dot{c}_{e}^{(i)}(t)=-\frac{\gamma_0}{2} c_e^{(i)}(t) 
-  \sum_n   \sum_{j=1}^{N_{ind}} \int_0^{+\infty}\mathrm{d}\omega e^{-i(\omega-\omega_0)t} \left[K_{n-,\omega}^{(i,j)}  \mathbf{c}_{j,\omega}^{(n-)}(t)+K_{n-,\omega}^{(i,j)}  \mathbf{c}_{j,\omega}^{(n+)}(t)\right]
\\
\label{eq:Cn_dynNe}
&& \dot{\mathbf{c}}_{j,\omega}^{(n-)}(t)= \sum_{i=1}^{N_e} K_{n+,\omega}^{(i,j)\star} e^{i(\omega-\omega_0)t}c_e^{(i)}(t)
\; ;\;
\dot{\mathbf{c}}_{j,\omega}^{(n+)}(t)= \sum_{i=1}^{N_e} K_{n-,\omega}^{(i,j)\star} e^{i(\omega-\omega_0)t}c_e^{(i)}(t)
\end{eqnarray}
\end{widetext}

For a system initially excited by the fundamental guided SPP$_0^+$ and all emitters in their ground state, the formal solutions of equations (\ref{eq:Cn_dynNe}) can be written as 
\begin{eqnarray}
&&\mathbf{c}_{j,\omega}^{(0-)}(t)=\sum_{i=1}^{N_e} K_{0-,\omega}^{(i,j)\star}  \int_0^t e^{i(\omega-\omega_0)t'}c_e^{(i)}(t')dt' \\
\nonumber 
&&\mathbf{c}_{j,\omega}^{(0+)}(t)=\mathbf{c}_{j,\omega}^{(0+)}(0)+\sum_{i=1}^{N_e}K_{0+,\omega}^{(i,j)\star}   \int_0^t   e^{i(\omega-\omega_0)t'}c_e^{(i)}(t')dt' \\
\nonumber 
&&\mathbf{c}_{j,\omega}^{(n\pm)}(t)=\sum_{i=1}^{N_e}  K_{n\pm,\omega}^{(i,j)\star} \int_0^t  e^{i(\omega-\omega_0)t'}c_e^{(i)}(t')dt' 
\end{eqnarray}
We assume that the waveguide is initially excited in the  fundamental SPP$_0^+$ guided mode propagating along the nanowire. The initial condition for the mode amplitude at the positions $\mathbf{r}_i=(\mathbf{\rho},z_i)$ is
\begin{eqnarray}
\alpha_{\omega,i}^{(0+)}(t=0)  =A(\omega)e^{-z_i/(2L_{spp}(\omega))}e^{ik_{spp}(\omega)z_i}
\label{eq:SPPinit}
\end{eqnarray}
for a guided mode in phase with the first atom of the chain and $\mathbf{c}_{\omega,j}^{(0+)}(0)$ follows from Eq. \ref{eq:cj0}
\begin{eqnarray}
\mathbf{c}_{j,\omega}^{(0+)}(t=0) =\frac{1}{\sqrt{\lambda_{\omega,0+}^j}}\sum_{i=1}^{N_e} T_{\omega,0+}^{ij} \alpha_{\omega,i}^{(0+)}(t=0) 
\label{eq:SPPinit2}
\end{eqnarray}
$A(\omega)$ is the frequency profile ({\it e.g} a gaussian profile centered at $\omega_0$) and we take into account the dispersion on the mode propagation constant $k_{spp}(\omega)$ and propagation length $L_{spp}(\omega)$. 

\subsection{Stationnary limits}
The stationnary limits of the reflected and transmitted SPP modes correspond to the long time limit $t \rightarrow \infty$ \cite{Chen-Koenderink:11,Greenberg:24}
\begin{eqnarray}
\nonumber 
&&\mathbf{c}_{j,\omega}^{(0+)}(t\rightarrow \infty)=\mathbf{c}_{j,\omega}^{(0+)}(0)+\sum_{i=1}^{N_e} K_{0+,\omega}^{(i,j)\star}   C_e^{(i)}(\omega) \\
\nonumber 
&&\mathbf{c}_{j,\omega}^{(n\pm)}(t\rightarrow \infty)=\sum_{i=1}^{N_e}  K_{n\pm,\omega}^{(i,j)\star}    C_e^{(i)}(\omega)
\end{eqnarray}
The Laplace transform of the dynamical equations (\ref{eq:Ce_dynNe},\ref{eq:Cn_dynNe}) leads to 
\begin{eqnarray}
\nonumber
&&i(\omega_0-\nu)C_e^{(i)}(\nu)=-\frac{\gamma_0}{2} C_e^{(i)}(\nu) 
\\
\nonumber
&&
-\sum_n   \sum_{j=1}^{N_{ind}}  \int_0^{+\infty}\mathrm{d}\omega  \left[ K_{n-,\omega}^{(i,j)} C_{j,\omega}^{n-}(\nu)
+K_{n+,\omega}^{(i,j)}C_{j,\omega}^{n+}(\nu)\right]\\
\nonumber
&&i(\omega-\nu)C_{j,\omega}^{0+}(\nu)-c_{j,\omega}^{0+}(0)= \sum_{i=1}^{N_e}  K_{0+,\omega}^{(i,j)\star} C_e^{(i)}(\nu)\\
&&i(\omega-\nu)C_{j,\omega}^{0-}(\nu)=\sum_{i=1}^{N_e}  K_{0-,\omega}^{(i,j)\star}  C_e^{(i)}(\nu)\\
\nonumber
&&i(\omega-\nu)C_{j,\omega}^{n\pm}(\nu)=\sum_{i=1}^{N_e}  K_{n\pm,\omega}^{(i,j)\star}   C_e^{(i)}(\nu) \;, j, n\ge 1
\end{eqnarray}
The $n^{th}$ coefficient follows 
\begin{eqnarray}
\nonumber 
C_{j,\omega}^{n\pm}(\nu)&=&\zeta(\nu-\omega)\sum_{i=1}^{N_e}  K_{n\pm,\omega}^{(i,j)\star}  C_e^{(i)}(\nu)  \;, n\ge 1\\
\nonumber \\
C_{j,\omega}^{0-}(\nu)&=&\zeta(\nu-\omega) \sum_{i=1}^{N_e}K_{0-,\omega}^{(i,j)\star}C_e^{(i)}(\nu) \\
C_{j,\omega}^{0+}(\nu)&=&\zeta(\nu-\omega) \left[c_{j,\omega}^{0+}(0)+ \sum_{i=1}^{N_e}K_{0+,\omega}^{(i,j)\star}C_e^{(i)}(\nu) \right] 
\nonumber 
\end{eqnarray}

The emitter's coefficient are solutions of the system of equations Eq. \eqref{eq:NemMatrix}
\begin{widetext}
\begin{eqnarray}
\nonumber
&&\left[i\left(\nu-\omega_0-\sum_n  \sum_{j=1}^{N_{ind}} \Omega_n^{(i,j)} \right)-\frac{\gamma_0}{2} -\sum_n  \sum_{j=1}^{N_{ind}} \frac{\Gamma_n^{(i,j)}}{2} \right]C_e^{(i)}(\nu)
\\ \nonumber && \hspace{1cm}
\approx
2\pi \sum_{j=1}^{N_{ind}} K_{0+,\nu}^{(i,j)}  c_{j,\omega}^{0+}(0)
+\pi \sum_n   \sum_{j=1}^{N_{ind}} \sum_{\substack{k=1 \\ k\neq i}}^{N_e}  \left[K_{n-,\nu}^{(i,j)}  K_{n-,\nu}^{(k,j)\star}+K_{n+,\nu}^{(i,j)}  K_{n+,\nu}^{(k,j)\star}\right]   C_e^{(k)}(\nu)
\\ \nonumber &&
\\
\nonumber
&&\Omega_n^{(i,j)} = \int_0^{+\infty}\mathrm{d \omega}\frac{\vert K_{n-,\omega}^{(i,j)}\vert^2}{\omega-\nu} + \int_0^{+\infty}\mathrm{d \omega}\frac{\vert K_{n+,\omega}^{(i,j)}\vert^2}{\omega-\nu} \\
&&\Gamma_n^{(i,j)} = 2\pi \vert K_{n-,\omega}^{(i,j)}\vert^2+2\pi \vert K_{n+,\omega}^{(i,j)}\vert^2
\nonumber
\end{eqnarray}
\end{widetext}

\subsection{Dipole-dipole shift}
\begin{figure}[h!]
\includegraphics[width=7cm]{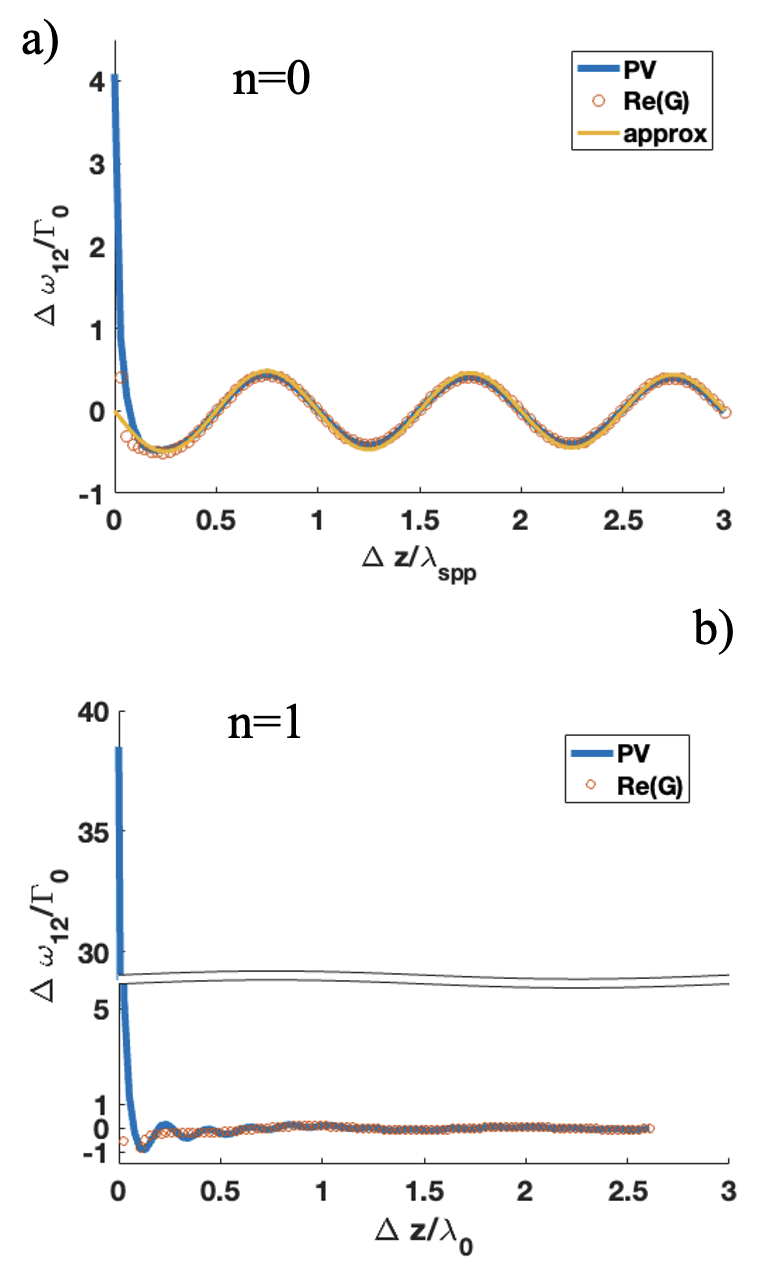}
\caption{Lamb shift calculated with exact principal value expression ('PV'), extending the integration to $-\infty$ ('real(G)') or approximated for a lossy propagative SPP ('approx'). a) Fundamental SPP$_0$ mode. b) n=1. In a) the distance is normalized with respect to the SPP wavelength $\lambda_{spp}=\lambda_0/n_{spp}$ whereas is it normalized with respect to the emission wavelength $\lambda_0=2\pi c/\omega_{0}$ in b). }
\label{fig:Lamb}   
\end{figure}

The expression of the amplitude entails dipole-dipole shifts. For instance, for two emitters, 

\begin{eqnarray*}
&&\Omega_n^{(i,j)} = \frac{1}{\hbar \pi \epsilon_0} {\displaystyle {\cal P} \int_0^{+\infty } \mathrm{d}\omega }\frac{\omega^2}{c^2}\frac{Im[{\mathbf d^{(i)}}\cdot {\mathbf G}_{n\pm}({\mathbf r}_i,{\mathbf r}_j,\omega) \cdot {\mathbf d^{(j)}}]}{\omega-\nu}
\end{eqnarray*}
They can be numerically computed using the property 
\begin{eqnarray}
\nonumber
&&{\displaystyle {\cal P} \int_a^{b} \mathrm{d}\omega}\dfrac{f(\omega)}{\omega-\omega_0} =\int_a^{b} \mathrm{d}\omega\dfrac{f(\omega)-f(\omega_0)}{\omega-\omega_0} +f(\omega_0)\ln {\left \vert \dfrac{b-\omega_0}{\omega_0-a} \right \vert }\\
&&{\displaystyle {\cal P} \int_0^{+\infty } \mathrm{d}\omega}\dfrac{f(\omega)}{\omega-\omega_0}= \int_0^{\omega_0} \mathrm{d}\omega\dfrac{f(\omega)-f(\omega_0)}{\omega-\omega_0} \\
\nonumber
&&\hspace{1cm}+\int_{\omega_0}^{2\omega_0} \mathrm{d}\omega\dfrac{f(\omega)-f(\omega_0)}{\omega-\omega_0}+ \int_{2\omega_0}^{+\infty} \mathrm{d}\omega\dfrac{f(\omega)}{\omega-\omega_0}
\end{eqnarray}
Some approximate expression can be also obtained. Indeed, if we extend the integral to $-\infty$ and use Kramers-Kronig relation, we obtain \cite{Fleischhauer:2011}
\begin{eqnarray}
\nonumber
\Omega_n^{(i,j)} &\approx& \frac{1}{\hbar \pi \epsilon_0} {\displaystyle {\cal P} \int_{-\infty }^{+\infty } \mathrm{d}\omega }\frac{\omega^2}{c^2}\frac{Im[{\mathbf d^{(i)}}\cdot {\mathbf G}_{n}({\mathbf r}_i,{\mathbf r}_j,\omega) \cdot {\mathbf d^{(j)}}]}{\omega-\nu} \\
&\approx&\frac{1}{\hbar \pi \epsilon_0}\frac{\nu^2}{c^2}Re[{\mathbf d^{(i)}}\cdot {\mathbf G}_{n}({\mathbf r}_i,{\mathbf r}_j,\nu) \cdot {\mathbf d^{(j)}}]
\end{eqnarray}
In addition, above the cut-off, this expression strongly simplifies \cite{Fleischhauer:2011}. For instance for the fundamental mode, we introduce the mode propagation constant $k_{spp}$ and propagation length $L_{spp}$ to obtain
\begin{eqnarray*}
\frac{\Omega_n^{(i,j)}}{\Gamma_0} &\approx& -\frac{1}{2}\sin(k_{spp}\Delta z)e^{-\Delta z/(2Lspp)}
\end{eqnarray*}
where $\Gamma_0$ refers to the coupling  rate of a single emitter it SPP$_0$. We plot in Fig. \ref{fig:Lamb}a, the dipole-dipole shift for the fundamental mode SPP$_0$. Apart from short distances, we observe that the approximate expressions correctly describe the dipole-dipole shift. However, at small distances or for a single emitter, $\Omega_0 \simeq 4\Gamma_0\simeq 60\gamma_0$ (the Purcell factor is  $\Gamma_0/\gamma_0 \simeq 15$) that could lead to a large shift compared to the free space emission angular frequency. 

Fig. \ref{fig:Lamb}b) presents the dipole-dipole shift for $n=1$. Since the nanowire diameter is below the cut off, one cannot use the modal approximation. The extension of the integral to $-\infty$ is not valid anymore but we observe that the contribution to the dipole-dipole shift decays fast as a function of distance. We used exact expressions to compute the figure in the main text but approximate expression could be used also at large distances. 
We observe that the Lamb shift for a single emitter is almost constant over the frequency range (not shown) so that we implicitly include it into  $\Omega_0 \approx \omega_0+\sum_n \Omega_n(\nu)$. However, actually the dip (peak) observed in the transmittivity(reflectivity) is shifted with respect to the isolated emitter due to the dipole-dipole coupling.

\section{{Limit in the sense of distributions expressed with Laplace transforms}}
\label{sect:DistribLimit}
We consider functions of two variables $f(\omega,t)$ and we define two types of limits. 
\begin{itemize}
\item[$\bullet$] The {\it regular function limit} is defined as
$\lim_{t\to\infty} f(\omega,t)  = F(\omega)$
if the limit exists pointwise with $F(\omega)$ a regular function.
\item[$\bullet$] The  { \it limit in the sense of distributions}, denoted by $d\lim_{t\to\infty} f_\omega(t)$, is defined, if it exists, as the unique distribution such that for all test  functions $\varphi(\omega)\in{\mathcal  D}$ \cite[p.14,98]{Strichartz-book-guide-distributions-Fourier}  

\begin{align}
 \int_{0}^\infty d\omega~ \left[ d\lim_{t\to\infty} f_\omega(t)\right] \varphi(\omega) 
&=\lim_{t\to\infty}  \int_{0}^\infty d\omega~ f_\omega(t) \varphi(\omega).
\end{align}
\\ \\
\end{itemize}

{\bf Theorem:} We consider functions $f_\omega(t)$ that are bounded in $\omega$ and bounded in $t$ for $t$ near infinity.
Then the limit in the sense of distributions $d\lim_{t\to\infty} f_\omega(t)$  exists and it can be expressed in terms o the Laplace transformation as
\begin{align}
d\lim_{t\to\infty} f_\omega(t) &=  \lim_{p\to0^+} p \hat f_\omega(p) \\
\text{with~}  \hat f_\omega(p) &:=  \mathcal{L}[f_\omega] (\omega,p) =\int_0^\infty dt~e^{-pt} f_\omega(t),
\end{align}
i.e. for any test function $\varphi(\omega)\in \mathcal{D}$
\begin{align}
 \int_{0}^\infty d\omega~ \left[ d\lim_{t\to\infty} f_\omega(t)\right] \varphi(\omega) 
&=\lim_{t\to\infty}  \int_{0}^\infty d\omega~ f_\omega(t) \varphi(\omega) \\
&=  \int_{0}^\infty d\omega~ \lim_{p\to0^+} p \hat f_\omega(p)  \varphi(\omega). 
\end{align}
{\bf Proof:}
\begin{align}
\nonumber
\int_{0}^\infty d\omega~& \lim_{p\to0^+} p \hat f_\omega(p)  \varphi(\omega) \\
 &= \int_{0}^\infty d\omega~ \lim_{p\to0^+} p  \int_0^\infty dt~e^{-pt} f_\omega(t)  \varphi(\omega) \\
  \label{change-var}
  (t':= pt), \;&=\int_{0}^\infty d\omega~ \lim_{p\to0^+}   \int_0^\infty dt'~e^{-t'} f_\omega({t'}/{p})  \varphi(\omega), \\
 \label{dom-conv-1}
  &= \lim_{p\to0^+} \int_{0}^\infty d\omega~  \int_0^\infty dt'~e^{-t'} f_\omega({t'}/{p})  \varphi(\omega) \\
  \label{Fubini-Tonelli}
  &= \lim_{p\to0^+} \int_0^\infty dt' \int_{0}^\infty d\omega~ ~e^{-t'} f_\omega({t'}/{p})  \varphi(\omega) \\
   \label{dom-conv-2}
  &= \int_0^\infty dt'~ e^{-t'}  \lim_{p\to0^+} \int_{0}^\infty d\omega~ f_\omega({t'}/{p})  \varphi(\omega) \\
  \label{lim-s}
   (s:=t'/p),\; &= \int_0^\infty dt'~ e^{-t'}  \lim_{s\to \infty} \int_{0}^\infty d\omega~ f_\omega(s)  \varphi(\omega),\\
      &=  \lim_{s\to \infty} \int_{0}^\infty d\omega~ f_\omega(s)  \varphi(\omega).
\end{align}
In \eqref{dom-conv-1} and in \eqref{dom-conv-2} we have used the dominated convergence theorem, and in 
\eqref{Fubini-Tonelli} the Fubini-Tonelli theorem.
In \eqref{lim-s} we have used that 
\begin{align}
 \lim_{p\to0^+} \int_{0}^\infty d\omega~ f_\omega({t'}/{p})  \varphi(\omega) 
 = \lim_{s\to \infty} \int_{0}^\infty d\omega~ f_\omega(s)  \varphi(\omega)
\end{align}
is independent of $t'$.

We consider differential equations of the general form
$\dot c = F(t).$ 
Then, the Laplace transform 
\begin{align}
 \tilde C(p)= \int_0^\infty dt~ e^{-pt} c(t)
\end{align}
satisfies
\begin{align}
p \tilde C(p) - c(t=0) = \tilde F(p),
\end{align}
and the solution is
\begin{align}
 \tilde C(p)  =  \frac{\tilde F(p) + c(t=0)}{p}
\end{align}
The limit $t\to\infty$ in the sense of distributions is given by \cite[p.24]{Dyke-book-Laplace-transforms-2014} 
\begin{align}
d\lim_{t\to\infty} c(t)  =  \lim_{p\to0^+} p~ \tilde C (p)
=  \lim_{p\to0^+} \tilde F(p) + c(t=0).
\end{align}

\subsection{Derivation of Eq. (\ref{eq:cn_Inftyc})}
Applying this formula to the differential equation Eq. (\ref{Ce_dynbc}), i.e. with $F(t) = K^{n*}_\omega e^{i(\omega-\omega_0)t} c_e(t)$ leads to the result
\begin{align}
\nonumber
d\lim_{t\to\infty} c^{n\pm}_\omega(t)  &=  \lim_{p\to0^+} p~ \tilde c^{\kern1pt n\pm}_\omega (p)
=  \lim_{p\to0^+} p  \int_0^\infty dt~ e^{-pt} c^{n\pm}_\omega(t) \\
\label{eq:dcn_Infty}
&= c^{n\pm}_\omega(t=0)  \\
\nonumber
&+\hspace{1cm} K_\omega^{n*} \lim_{p\to 0^+} p \int_0^\infty dt~ e^{-pt} e^{i(\omega-\omega_0)} c_e(t)  \\
& \equiv  c^{n\pm}_\omega(t=0) + K_\omega^{n*}  \lim_{p\to 0^+}  \check{C}_e(-p+i(\omega-\omega_0)). 
\nonumber
\end{align}

Where we have defined the following function 
\begin{align*}
\check{C}_e(-p+i(\omega-\omega_0))=\int_0^\infty dt~ e^{-pt}~ e^{i(\omega-\omega_0)t} c_e(t).
\end{align*}
as the Laplace transform of $e^{i(\omega-\omega_0)t} c^{n\pm}_\omega(t)$. 
Equation (\ref{eq:dcn_Infty}) is abusively simplified to Eq. (\ref{eq:cn_Inftya}-\ref{eq:cn_Inftyc}) in the main text, with initial conditions $c^{n\pm}_\omega(t=0)\; , n\ge 1$ and $c^{0-}_\omega(t=0)=0$. 

\subsection{Derivation of Eq. (\ref{C_LapA})}
In the next steps we will determine $ \check{C}_e(-p+i(\omega-\omega_0))$. 
First, we determine the relation between $\check{C}_e$ and $\check{C}^{n\pm}_\omega$. \\
We apply $\int_0^\infty dt~ e^{-pt} e^{i(\omega-\omega_0)t}$ to the differential equation Eq. (\ref{Ce_dyna}):
\begin{widetext}
\begin{align}
\int_0^\infty dt~ e^{-pt} e^{i(\omega-\omega_0)}\dot c_e(t) 
&= -\frac{\gamma_0}{2} \int_0^\infty dt~ e^{-pt} e^{i(\omega-\omega_0)} c_e(t)  \\ \nonumber & \hspace{1cm}
 - \sum_n\int_0^\infty d\omega'~K^{n}_{\omega'} \int_0^\infty dt~ e^{-pt} e^{i(\omega-\omega_0)t}  e^{-i(\omega'-\omega_0)t} \left[  c^{n+}_{\omega'}(t)+ c^{n-}_{\omega'}(t) \right]  \\
 \nonumber
&=  -\frac{\gamma_0}{2} \check{C}_e(-p+i(\omega-\omega_0))
 - \sum_n\int_0^\infty d\omega'~K^{n}_{\omega'} \int_0^\infty dt~ e^{-pt} e^{-i(\omega'-\omega)t}  \left[  c^{n+}_{\omega'}(t)+ c^{n-}_{\omega'}(t) \right] \\
 &=  -\frac{\gamma_0}{2} \check{C}_e(-p+i(\omega-\omega_0))
 - \sum_n\int_0^\infty d\omega'~K^{n}_{\omega'}  \sum_\pm \check{C}^{n\pm}_{\omega'}(-p-i(\omega'-\omega)).
 \nonumber
\end{align}
\end{widetext}

Thanks to integration by parts, the left hand side term can be written as
\begin{widetext}
\begin{align*}
\int_0^\infty dt~ e^{-pt} e^{i(\omega-\omega_0)}\dot c_e(t)
 &=-(-p+i(\omega-\omega_0))\int_0^\infty dt~ e^{-pt} e^{i(\omega-\omega_0)} c_e(t) 
+\left[e^{-pt} e^{i(\omega-\omega_0)} c_e(t)  \right]_{t=0}^{t=\infty}  \\
 &=(p-i(\omega-\omega_0))    \check{C}_e(-p+i(\omega-\omega_0)) - c_e(t=0),
\end{align*}
\end{widetext}
and thus,
\begin{eqnarray*}
&&\left[ p-i(\omega-\omega_0) +\frac{\gamma_0}{2}   \right] \check{C}_e(-p+i(\omega-\omega_0)) \\
\nonumber 
 &&= c_e(t=0)- \sum_n\int_0^\infty d\omega'~K^{n}_{\omega'}   \sum_\pm \check{C}^{n\pm}_{\omega'}(-p-i(\omega'-\omega)).
 \label{Cce}
\end{eqnarray*}
summarized by Eq. (\ref{C_LapA}) in the main text with the limit $p\rightarrow0^+$.

In the next step, we express $\check{C}^{n\pm}_{\omega'}$ in terms of $\check{C}_e$ in order to obtain an equation in which $\check{C}_e$  is the only unknown.
Applying $\int_0^\infty dt~ e^{-pt} e^{-i(\omega'-\omega)t}$ to the differential equation Eq. (\ref{Ce_dynbc}), we obtain
\begin{align}
 \nonumber
&\int_0^\infty dt~ e^{-pt} e^{-i(\omega'-\omega)t}\frac{d}{dt} c^{n\pm}_{\omega'}(t)  \\
 \nonumber
&= K^{n*}_{\omega'}\int_0^\infty dt~ e^{-pt} e^{-i(\omega'-\omega)t} e^{i(\omega'-\omega_0)t} c_e(t)\\
 \nonumber
&= K^{n*}_{\omega'}\int_0^\infty dt~ e^{-pt} e^{i(\omega- \omega_0)t} c_e(t) \\
&\equiv K^{n*}_{\omega'} \check{C}_e(-p+i(\omega-\omega_0)).
\end{align}
The lefthand side can be written as
\begin{align}
&\int_0^\infty dt~ e^{-pt} e^{-i(\omega'-\omega)t}\frac{d}{dt} c^{n\pm}_{\omega'}(t)  \\
 \nonumber
&= \left[ p+i(\omega'-\omega)  \right] \int_0^\infty dt~e^{-pt} e^{-i(\omega'-\omega)t}c^{n\pm}_{\omega'}(t)  \\
 \nonumber
&+\left[e^{-pt} e^{-i(\omega'-\omega)t}c^{n\pm}_{\omega'}(t)  \right]_{t=0}^{t=\infty} \\
 \nonumber
&= \left[ p+i(\omega'-\omega)  \right] \check{C}^{n\pm}_{\omega'}(-p-i(\omega'-\omega))  
-c^{n\pm}_{\omega'}(t=0),
\end{align}
and thus,
\begin{align}
& \left[ p+i(\omega'-\omega)  \right] \check{C}^{n\pm}_{\omega'}(-p-i(\omega'-\omega))\\
& \nonumber  
= c^{n\pm}_{\omega'}(t=0)+ K^{n*}_{\omega'} \check{C}_e(-p+i(\omega-\omega_0)),
\end{align}
which leads to
\begin{eqnarray} 
\label{C-Ce}
&&\check{C}^{n\pm}_{\omega'}(-p-i(\omega'-\omega))\\
\nonumber
&&=\dfrac{c^{n\pm}_{\omega'}(t=0)+ K^{n*}_{\omega'} \check{C}_e(-p+i(\omega-\omega_0))}
{p+i(\omega'-\omega)}.
\end{eqnarray}
summarized by Eq. \eqref{C_LapA} in the main text for $p\rightarrow 0$.

\subsection{Derivation of Eq. ( \ref{eq:Ce_nu})}
 Inserting \eqref{C-Ce} into \eqref{Cce} we obtain 
\begin{widetext}
\begin{align}
 \nonumber
 &\left[ p-i(\omega-\omega_0) +\frac{\gamma_0}{2}   \right] \check{C}_e(-p+i(\omega-\omega_0))  \\
  \nonumber
& = c_e(t=0)
 - \sum_n\int_0^\infty d\omega'~K^{n}_{\omega'}   \sum_\pm \frac{c^{n\pm}_{\omega'}(t=0)+ K^{n*}_{\omega'} \check{C}_e(-p+i(\omega-\omega_0))} {p+i(\omega'-\omega)} \\
  \nonumber
 & = c_e(t=0)- \sum_n\int_0^\infty d\omega'~ \frac{K^{n}_{\omega'}   \sum_\pm c^{n\pm}_{\omega'}(t=0)  }{p+i(\omega'-\omega)}
   -2 \sum_n\int_0^\infty d\omega'~\frac{|K^{n}_{\omega'}|^2  \check{C}_e(-p+i(\omega-\omega_0))} {p+i(\omega'-\omega)} \\
    & =  c_e(t=0)    
    - \sum_n\int_0^\infty d\omega'~ \frac{K^n_{\omega'}\sum_\pm c^{n\pm}_{\omega'}(t=0) }{p+i(\omega'-\omega)} 
  - \check{C}_e(-p+i(\omega-\omega_0)) ~2  \sum_n\int_0^\infty d\omega'~\frac{|K^{n}_{\omega'}|^2  } {p+i(\omega'-\omega)}, 
\end{align}
\end{widetext}

and further, using the fact that the initial coefficients are all zero except $c^{0+}_{\omega'}(t=0)$, 
\begin{widetext}
\begin{align}
 \nonumber
 &\left[ p-i(\omega-\omega_0) +\frac{\gamma_0}{2}  +2  \sum_n\int_0^\infty d\omega'~\frac{|K^{n}_{\omega'}|^2  }
  {p+i(\omega'-\omega)} \right] \check{C}_e(-p+i(\omega-\omega_0)) 
  = -\int_0^\infty d\omega'~ \frac{K^0_{\omega'}c^{0+}_{\omega'}(t=0) }{p+i(\omega'-\omega)}\\
 \nonumber
 &\left[ p-i(\omega-\omega_0) +\frac{\gamma_0}{2}  +2  \sum_n\int_0^\infty d\omega'~\frac{i|K^{n}_{\omega'}|^2  }
  {\omega-\omega'+ip} \right] \check{C}_e(-p+i(\omega-\omega_0)) 
 = - \int_0^\infty d\omega'~ \frac{iK^0_{\omega'}c^{0+}_{\omega'}(t=0) }{\omega-\omega'+ip}
\\
 \label{Ce-explicit}
 &\check{C}_e(-p+i(\omega-\omega_0))  
 = \frac { - \int_0^\infty d\omega'~ \dfrac{iK^0_{\omega'}c^{0+}_{\omega'}(t=0) }{\omega-\omega'+ip} }
    { p-i(\omega-\omega_0) +\frac{\gamma_0}{2}  +2  \sum_n\int_0^\infty d\omega'~\dfrac{i|K^{n}_{\omega'}|^2  }
  {\omega-\omega'+ip} }.
\end{align}
\end{widetext}
summarized by Eq. (\ref{eq:Ce_nu}) in the main text for $p\rightarrow 0$.

\subsection{Derivation of Eq. ( \ref{eq:C0Infty})}
Starting from \eqref{Ce-explicit}, we obtain
\begin{eqnarray}
&& \lim_{p\to 0^+}  \check{C}_e(-p+i(\omega-\omega_0))  
   \\
   \nonumber 
  &&= \frac { - \int_0^\infty d\omega'~ \frac{iK^0_{\omega'}c^{0+}_{\omega'}(t=0) }{\omega-\omega'+i0^+} }
    { -i(\omega-\omega_0) +\frac{\gamma_0}{2}  +2  \sum_n\int_0^\infty d\omega'~\frac{i |K^{n}_{\omega'}|^2  }
  {\omega-\omega'+i0^+} },
\end{eqnarray}
and using 
\begin{flalign*}
& \zeta(\omega-\omega') = \frac{i  } {\omega-\omega'+i0^+} =\pi\delta(\omega-\omega')+i\mathcal{P}\left(  \frac{1}{\omega-\omega'} \right),   
\end{flalign*}
we can write
\begin{eqnarray}
&& \lim_{p\to 0^+}  \check{C}_e(-p+i(\omega-\omega_0))  \\
\nonumber 
&&= \frac{  \int_0^\infty d\omega'~ \zeta(\omega-\omega')~K^{0}_{\omega'}~c^{0+}_{\omega'}(t=0) }  { i(\omega-\omega_0) - \left[\frac{\gamma_0}{2}  +  \sum_n \Gamma_n(\omega)/2\right] -  i\sum_n \Omega_n(\omega) } \\
\nonumber 
      &&= \frac {  \pi K^{0}_{\omega}~c^{0+}_{\omega}(t=0) 
      +  i\mathcal{P}\int_0^\infty d\omega'~\frac{K^{0}_{\omega'}~c^{0+}_{\omega'}(t=0)} {\omega-\omega' }}   { i(\omega-\omega_0) -\left[\frac{\gamma_0}{2}  +  \sum_n \Gamma_n(\omega)/2 \right] -  i\sum_n \Omega_n(\omega) }.        
\end{eqnarray}
These results, together with  Eq. \eqref{eq:dcn_Infty}  imply the following main result
\\ 
\begin{eqnarray}
\nonumber 
&&d\lim_{t\to\infty} c^{n\pm}_\omega(t)  
:=  \lim_{p\to0^+} p  \int_0^\infty dt~ e^{-pt} c^{n\pm}_\omega(t) \\
\nonumber 
&& \equiv  c^{n\pm}_\omega(t=0) + K_\omega^{n*}  \lim_{p\to 0^+}  \check{C}_e(-p+i(\omega-\omega_0)) \\
&& = c^{n\pm}_\omega(t=0) \\
\nonumber 
&&+ K_\omega^{n*}  
\frac {  \pi K^{0}_{\omega}~c^{0+}_{\omega}(t=0) 
      +  i\mathcal{P}\int_0^\infty d\omega'~\frac{K^{0}_{\omega'}~c^{0+}_{\omega'}(t=0)} {\omega-\omega' }}  
       { i(\omega-\omega_0) -\left[\frac{\gamma_0}{2}  +  \sum_n \Gamma_n(\omega)/2 \right]-  i\sum_n \Omega_n(\omega) }
      \label{Ce-final}
\end{eqnarray}
and specifically for $n=0$ and  $n\geq 1$ ; 
\begin{widetext}
\begin{eqnarray}
&&d\lim_{t\to\infty} c^{0+}_\omega(t) 
 = c^{0+}_\omega(t=0) + \dfrac { \Gamma_0(\omega)/4~c^{0+}_{\omega}(t=0) 
 +  i K_\omega^{0*} \mathcal{P}\int_0^\infty d\omega'~\frac{K^{0}_{\omega'}~c^{0+}_{\omega'}(t=0)} {\omega-\omega' }}   { i(\omega-\omega_0) -\left[\frac{\gamma_0}{2}  +  \sum_n \Gamma_n(\omega)/2 \right]-  i\sum_n \Omega_n(\omega) }\\
&&d\lim_{t\to\infty} c^{0-}_\omega(t)  
=\frac { \Gamma_0(\omega)/4~c^{0+}_{\omega}(t=0) 
      +  i K_\omega^{0*} \mathcal{P}\int_0^\infty d\omega'~\frac{K^{0}_{\omega'}~c^{0+}_{\omega'}(t=0)} {\omega-\omega' }}   { i(\omega-\omega_0) -\left[\frac{\gamma_0}{2}  +  \sum_n \Gamma_n(\omega)/2 \right]-  i\sum_n \Omega_n(\omega) }
\\
&&d\lim_{t\to\infty} c^{n\pm}_\omega(t) = K_\omega^{n*}  
\frac {  \pi K^{0}_{\omega}~c^{0+}_{\omega}(t=0) 
      +  i\mathcal{P}\int_0^\infty d\omega'~\frac{K^{0}_{\omega'}~c^{0+}_{\omega'}(t=0)} {\omega-\omega' }}   
      { i(\omega-\omega_0) -\left[\frac{\gamma_0}{2}  +  \sum_n \Gamma_n(\omega)/2 \right]-  i\sum_n \Omega_n(\omega) } \; (n\geq 1).
\end{eqnarray}
\end{widetext}
Multiplying numerators and denominators by $-i$ and regrouping real and imaginary parts in the denominators one can write
\begin{widetext}
\begin{eqnarray}
&&d\lim_{t\to\infty} c^{0+}_\omega(t)  
 = c^{0+}_\omega(t=0) + 
\frac { -i \Gamma_0(\omega)/4~c^{0+}_{\omega}(t=0) 
      +   K_\omega^{0*} \mathcal{P}\int_0^\infty d\omega'~\frac{K^{0}_{\omega'}~c^{0+}_{\omega'}(t=0)} {\omega-\omega' }}   { \omega-\omega_0-\sum_n \Omega_n(\omega)  +\frac{i}{2}  \left[ \gamma_0+\sum_n \Gamma_n(\omega) \right] }
      \label{c0+final}\\
&&d\lim_{t\to\infty} c^{0-}_\omega(t)  
=\frac { -i \Gamma_0(\omega)/4~c^{0+}_{\omega}(t=0) 
      +   K_\omega^{0*} \mathcal{P}\int_0^\infty d\omega'~\frac{K^{0}_{\omega'}~c^{0+}_{\omega'}(t=0)} {\omega-\omega' }}   { \omega-\omega_0-\sum_n \Omega_n(\omega)  +\frac{i}{2}  \left[ \gamma_0+\sum_n \Gamma_n(\omega) \right] }
      \label{c0-final}\\
&&d\lim_{t\to\infty} c^{n\pm}_\omega(t)  
= K_\omega^{n*}  
\frac { -i \pi K^{0}_{\omega}~c^{0+}_{\omega}(t=0) 
      +  \mathcal{P}\int_0^\infty d\omega'~\frac{K^{0}_{\omega'}~c^{0+}_{\omega'}(t=0)} {\omega-\omega' }}  
      { \omega-\omega_0-\sum_n \Omega_n(\omega)  +\frac{i}{2}  \left[ \gamma_0+\sum_n \Gamma_n(\omega) \right] } \; (n\geq 1).
      \label{cn-final}
\end{eqnarray}
\end{widetext}
In order to compare with the formula \eqref{eq:C0Infty} in the main text,  the expression \eqref{c0+final} can be written as
\begin{widetext}
\begin{eqnarray}
\nonumber
&&d\lim_{t\to\infty} c^{0+}_\omega(t)  
 = c^{0+}_\omega(t=0) \left[ 1+
\frac { -i \Gamma_0(\omega)/4}
 { \omega-\omega_0-\sum_n \Omega_n(\omega)  +\frac{i}{2}  \left[ \gamma_0+\sum_n \Gamma_n(\omega) \right] }   \right]
    + \frac {   K_\omega^{0*} \mathcal{P}\int_0^\infty d\omega'~\frac{K^{0}_{\omega'}~c^{0+}_{\omega'}(t=0)} {\omega-\omega' }}  
  { \omega-\omega_0-\sum_n \Omega_n(\omega)  +\frac{i}{2}  \left[ \gamma_0+\sum_n \Gamma_n(\omega) \right] }
      \label{c0+final-22}
\end{eqnarray}
\end{widetext}
and further, using
\begin{eqnarray*}
i\sum_n \Gamma_n(\omega)/2  -i \Gamma_0(\omega)/4
&= i \Gamma_0(\omega)/4 + i\sum_{n\geq1} \Gamma_n(\omega)/2
\end{eqnarray*}
we obtain the expression Eq. ( \ref{eq:C0Infty}) in the sense of distribution 
\newpage
\begin{widetext}
\begin{eqnarray}
\nonumber
&&d\lim_{t\to\infty} c^{0+}_\omega(t)  \\  \nonumber
&& = c^{0+}_\omega(t=0) \left[
\frac {{\omega-\omega_0-\sum_n \Omega_n(\omega)  +\frac{i}{2}  \left[ \gamma_0+ \Gamma_0(\omega)/2+\sum_{n\geq1} \Gamma_n(\omega) \right] }}
 { \omega-\omega_0-\sum_n \Omega_n(\omega)  +\frac{i}{2}  \left[ \gamma_0+\sum_n \Gamma_n(\omega) \right] }   \right]
    + \frac {   K_\omega^{0*} \mathcal{P}\int_0^\infty d\omega'~\frac{K^{0}_{\omega'}~c^{0+}_{\omega'}(t=0)} {\omega-\omega' }}  
  { \omega-\omega_0-\sum_n \Omega_n(\omega)  +\frac{i}{2}  \left[ \gamma_0+\sum_n \Gamma_n(\omega) \right] }.
      \label{c0+final-33}
\end{eqnarray}
\end{widetext}


\begin{acknowledgments}
European Regional Development Fund FEDER-FSE 2021-2027, Bourgogne-Franche-Comté ; French Investissements d’Avenir program EUR-EIPHI (17-EURE-0002); Calculations were performed on mesoBFC data center. 
\end{acknowledgments}

%

\end{document}